\pdfoutput=1
\documentclass[aps,nofootinbib,twocolumn,groupedaddress,floatfix,eqsecnum]{revtex4}
\usepackage{graphicx}
\usepackage{epsfig}
\usepackage{epstopdf}
\usepackage[mathscr]{eucal}
\usepackage{amsmath,amssymb,bm,graphics}
\usepackage[pdftex,letterpaper=true]{hyperref}

\newcommand{\mc}[1]{\mathcal{#1}}

\let\mathbf=\bm

\begin{document}
\def\Nfour{\mathcal N\,{=}\,4}
\def\Ntwo{\mathcal N\,{=}\,2}
\def\Nc{N_{\rm c}}
\def\Nf{N_{\rm f}}
\def\x{\mathbf x}
\def\q{\mathbf q}
\def\f{\mathbf f}
\def\v{\mathbf v}
\def\w{\omega}
\def\vs{v_{\rm s}}
\def\S{\mathcal S}
\def\half{{\textstyle \frac 12}}
\def\twothirds{{\textstyle \frac 23}}
\def\third{{\textstyle \frac 13}}
\def\t{\mathbf{t}}
\def\T{\mathcal {T}}
\def\O{\mathcal{O}}
\def\E{\mathcal{E}}
\def\p{\mathcal{P}}
\def\H{\mathcal{H}}
\def\uh{u_h}
\def\R{\ell}
\def\Ro{\chi}
\def\del{\nabla}
\def\eps{\hat \epsilon}
\def\nn{\nonumber}
\def\K{\mathcal K}
\def\inf{\epsilon}
\def\cs{c_{\rm s}}
\def\A{\mathcal{A}}
\def\e{{e}}
\def\r{{\xi}}
\def\x{{\mathbf x}}
\def\w{{w}}
\def\rr{{\xi}}
\def\G{\mathcal{G}}

\title
{Jets in strongly-coupled
$\mathcal N = 4$ super Yang-Mills theory }

\author{Paul~M.~Chesler\footnotemark}
\author{Kristan~Jensen\footnotemark}
\author{Andreas~Karch\footnotemark}

\affiliation
    {Department of Physics, University of Washington, Seattle, WA 98195, USA}

\date{\today}

\begin{abstract}
We study jets of massless particles in ${\cal N}=4$ super Yang-Mills
using the AdS/CFT correspondence both at zero and finite
temperature. We set up an initial state corresponding to a highly
energetic quark/anti-quark pair and follow its time evolution into
two jets. At finite temperature the jets stop after traveling a
finite distance, whereas at zero temperature they travel and spread
forever. We map out the corresponding baryon number charge density
and identify the generic late time behavior of the jets as well as
features that depend crucially on the initial conditions.
\end{abstract}


\maketitle
\iftrue
\def\thefootnote{\fnsymbol{footnote}}
\footnotetext[1]{Email: \tt pchesler@u.washington.edu}
\footnotetext[2]{Email: \tt kristanj@u.washington.edu}
\footnotetext[3]{Email: \tt karch@phys.washington.edu}
\def\thefootnote{\arabic{footnote}}
\fi

\section{Introduction}

In a hadronic collider such as the LHC, most interesting processes
involve final states with jets of hadrons or equivalently,
sprays of particles heading
in roughly the same direction within a certain opening angle.
The data available to define and reconstruct the jets are energy
depositions in the calorimeter and charged particle tracks.
Ideally one can use this experimental data to identify an
underlying
hard scattering event that involved QCD partons in the weakly
coupled high energy regimes or, alternatively, heavy
standard model particles such as top quarks or newly produced
non-standard model particles. Distinguishing the former from
the latter is of course crucial to our ability to find and
analyze new physics at the LHC.

Despite its obvious importance, several aspects of jet physics are
still in need of better understanding.  A large portion
of the current understanding of jets comes from perturbative
techniques.  For a recent review see Ref
\cite{Ellis:2007ib}. First, a jet has to be defined via one of
several algorithms, all of which have their own benefits and
problems. The properties of the jet then have to be compared to a
perturbative calculation of the hard scattering. The initial hard
scattering however has to be first processed by a parton shower
routine that evolves every hard parton into a jet of partons which
eventually hadronize into baryons and mesons. The distribution of
the latter can then be matched with the experimentally observed
data. This compartmentalization of the theoretical part into three
stages assumes that these processes, which are governed by physics
at very different energy scales, happen independently. This is the
statement of factorization. While resummed perturbation theory or
soft collinear effective theory give a theoretical handle on the
showering, Monte Carlo simulations based on models (and not
QCD) are often employed for this step. The hadronization is always
handled by phenomenological models. One question we will address in
this work is how the process of showering occurs in a particular
solvable toy model of strong coupling dynamics.
In particular, we study jets in large $N_c$, $SU(N_c)$
${\cal N}=4$
super-Yang-Mills (SYM) at strong 't~Hooft coupling $\lambda$ where
the process has a dual description in terms of a string falling
in a $5d$ geometry. This
may help us understand what aspects of showering are genuinely QCD
and may also lead to better models of showering.

A quite different experimental setting that prominently features
jets are heavy ion colliders, such as the currently operating RHIC
experiment as well as the heavy-ion runs that will take place part
of the time at the LHC.  The goal of these experiments is not so much
to find physics beyond the standard model but rather to explore the
phase structure of QCD at high temperatures.  At the
temperatures achieved in the RHIC collisions, quarks and gluons are
thought to be liberated although not free. Their collective dynamics
are well described by an almost perfect fluid \cite{Shuryak,Shuryak:2004cy}. 
Jets, which are
produced by hard collisions within the expanding fireball, serve as
a probe of the interior of the plasma. In this context one wants to
know the rate at which jets lose energy and broaden when traversing
the medium. These are strongly coupled non-equilibrium phenomena and
are therefore challenging for both the lattice and for diagrammatic approaches.
Again, it is helpful to have a solvable toy model.  We will study
jets of massless quarks traversing a finite temperature plasma in
the same solvable toy model, ${\cal N}=4$ SYM at strong coupling. 

Our approach, both at zero and finite temperature, is to prepare at
a state at time $t=0$ which is well described by a semi-classical
wave-function.  In the gravity dual, this corresponds to a classical
string moving in the dual geometry. At $t=0$ we specify the initial
configuration of the string and let it time evolve according to the
equations of motion that follow from the Nambu-Goto action.  The time
evolution of the string is dual to the evolution of baryon and energy
densities in the corresponding field theory state.  A similar approach 
has been used to study the properties of jets of massive matter in the same 
theory at zero temperature \cite{Lin:2007fa,Chernicoff:2008sa} and finite temperature 
\cite{Chernicoff:2008sa}.  In this work, we compute the evolution of baryon density 
at both zero and finite temperature for the case of massless quarks.
For concreteness, we consider states with two back
to back jets.  In particular, we consider string states where the 
string profile lies in one Minkowski spatial direction.
However our analysis and results for the baryon density are
easily generalized to the case where the string profile 
extends in more than one Minkowski spatial direction. 

For the classical description to be valid the string must carry large conserved
charges. In particular, the energy of the string scales as
$\sqrt{\lambda}$. At finite temperature, this setup is very well suited for the question of
studying the energy and momentum loss rates of an energetic
projectile traversing the medium. Our calculation generalizes the
calculation of the heavy quark energy loss rate of
\cite{Herzog:2006gh,Gubser:2006bz} to the case of massless quarks.
The additional complication is that for massless quarks the plasma
can transport the charge and so the initially localized charge
distribution eventually diffuses; the string falls. While we don't
have a conclusive answer on the loss rate and stopping distance yet,
our simulations indicate that a recent analysis of this process 
based purely on the analysis of geodesics \cite{Gubser:2008as} is
incomplete. One of the central results of our
numerical investigations is that the endpoint motion indeed very
quickly settles onto a lightlike geodesic for generic initial
conditions. However the relation between the energy of the quark and the
parameters specifying the geodesic can be more complicated than
presented in \cite{Gubser:2008as}.

At zero temperature our approach is new, even though the situation
is quite similar in spirit to the case of finite temperature. At time $t=0$ we prepare a
quark/antiquark pair with enough kinetic energy to climb out of each
others' potential well. We follow in
detail the evolution of the baryon number density and can in
principle do the same for energy densities. As each one of the energetic
projectiles moves through the vacuum, both densities spread so that
instead of two single well defined particles flying back-to-back at the speed of
light, we see narrow distributions of
baryon number density
within two cones around the directions of the original quark/antiquark
pair
--- a pair of jets. The dynamical mechanism underlying this jet
formation from the field
theory perspective seems to be the emission of a large number of
soft gluons and quarks.  This setup with an initial semi-classical
state whose subsequent evolution is studied mirrors closely the
usual factorization framework. An initial hard scattering event will
have to produce the $t=0$ string configuration. We do not specify
the dynamics that gave rise to the initial configuration. In any
case, this would have to be very different from QCD where the
initial hard event is perturbative. The subsequent showering is what
we study using the gravity dual. The conformal theory itself doesn't hadronize, 
but gravity duals for confining gauge theories are known. 
In these cases only the deep interior of the geometry is modified. This will not modify
the initial falling stage of the string motion. 
Factorization of the showering stage from
the hadronization stage is guaranteed by bulk locality in our setup.

This approach is rather different from the recent work in \cite{Hatta:2008tx}.  
There the authors study the evolution of a vector wavepacket as 
it falls into an AdS geometry.  The dynamics of the falling wavepacket are dual 
to the evolution of an $R$-charge jet in the same theory that we study, $\mc{N}=4$ 
SYM at strong coupling.  In contrast, we study jets of fundamental matter 
with large energies that scale like $\sqrt{\lambda}$.

An alternative approach to studying jets 
is to consider observables which are completely
self contained and do not refer to a particular initial state. The
best theoretically motivated observables are the energy
correlators introduced in
\cite{Basham:1977iq,Basham:1978bw,Basham:1978zq}. In a state
obtained from acting with a local current on the vacuum, the energy
density one-point function exhibits an ``antenna pattern'' where the
number of particles heading into a particular direction is
correlated with the polarization of the current, giving rise to an
intrinsically jet-like event. These antenna patterns have recently
been studied in \cite{Hofman:2008ar} for strongly coupled ${\cal
N}=4$ (in fact for any CFT) and it was found that in a theory with a
supergravity dual they are perfectly spherical and show no sign of
jet-like structure.%
\footnote
  {This result had been anticipated in
  \cite{Polchinski:2002jw, Strassler:2008bv,Lin:2007fa}.
  } 
A different question is to ask about the likelihood that an 
initial off shell photon would produce a quark/antiquark pair 
with a particular preferred axis. The result of 
\cite{Hofman:2008ar} shows that an off shell photon would 
produce quark/antiquark pairs with no preferred axis 
and on top of this no correlation between the direction 
of the quark and the antiquark. In contrast, we set up an 
initial state that has a preferred direction for both. We 
mostly focus on back-to-back jets, but as we show in Section 
\ref{discussion}, our methods apply for quarks 
and antiquarks heading in completely different directions. 
The results of \cite{Hofman:2008ar} then simply implies that 
the initial states we consider are not created by a local 
${\cal N}=4$ current operator. 
Again, we emphasize that this is not a concern 
for the issue of studing showering as the dynamics 
that give rise to the creation of a $q \bar q$ state via 
the decay of an energetic off-shell photon are qualitatively 
different in a strongly-coupled conformal theory than QCD, 
which is asymptotically free.

\begin{figure}[h]
\includegraphics[scale=0.7]{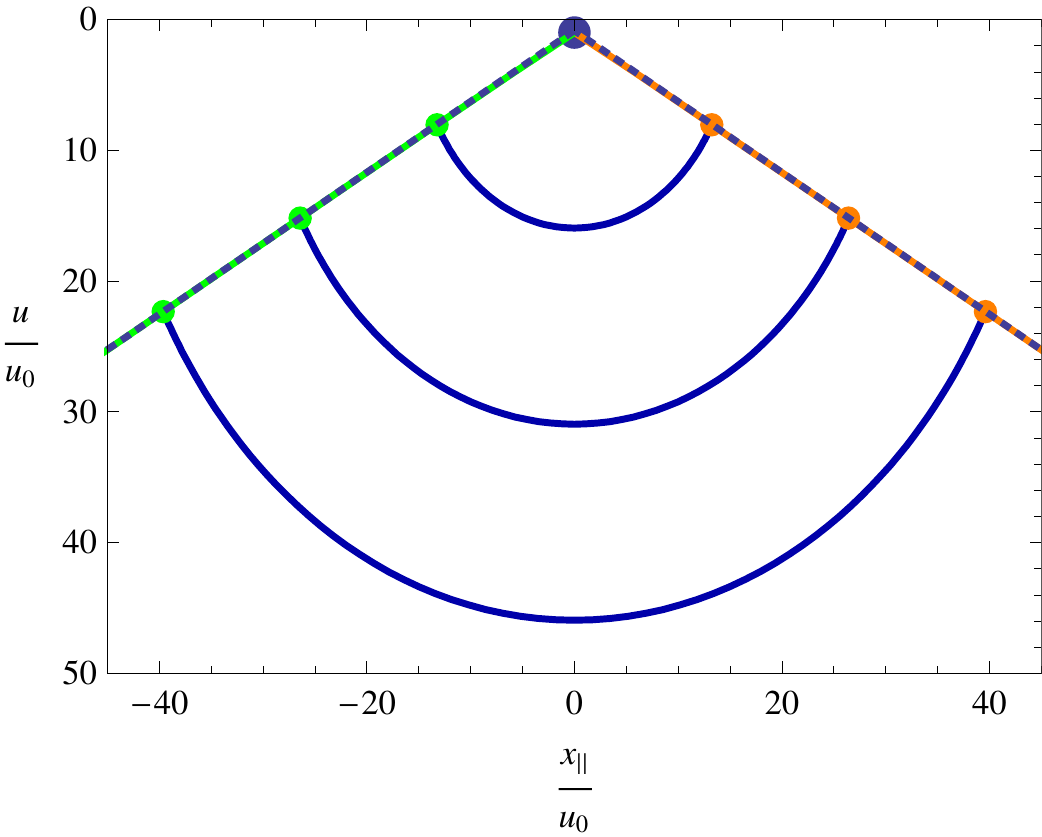}
\caption{\label{zeroTintroplot}
    A plot of a falling string at zero temperature.  The string is
    created at radial coordinate $u=u_0$ and expands as it falls.
    The endpoint trajectories, shown in the figure as solid
    green and orange lines, asymptotically approach light-light
    geodesics which are shown in the figure as dotted blue lines.
    }
\end{figure}

The framework that allows us to do these calculations is the AdS/CFT
correspondence \cite{Maldacena:1997re,Gubser:1998bc,Witten:1998qj}.
In the limit $\Nc \rightarrow \infty$ and $\lambda \gg 1$, ${\cal
N}=4$ SYM has a dual description in terms of type IIB supergravity on
an AdS$_5$ or AdS$_5$ black hole geometry times an internal
5-sphere. In addition, states with an energy scaling as
$\sqrt{\lambda}$ can be described by a classical Nambu-Goto string
moving in this spacetime, as is implicit in the calculation of Wilson
line expectation values of \cite{Maldacena:1998im,Rey:1998ik} and
was argued in detail in \cite{Gubser:2002tv}. Similarly, we consider
a state that consists of an open string which is created at a point
in space at time $t = 0$. Such initial conditions locally conserve
charge in the bulk and baryon number on the boundary.
As shown in Fig.~\ref{zeroTintroplot}, as time evolves the
string evolves from a point into an extended object and falls under
the influence of gravity.  The string's endpoints are charged under
a $U(1)$ gauge field $\mathcal A_M$. The simplest realization of
such a gauge field is the worldvolume gauge field of a D7 flavor
brane, which is spacetime filling from the 5d point of view and
wraps a 3-sphere in the internal $S^5$. According to
\cite{Karch:2002sh} the addition of such a flavor brane from the
field theory point of view corresponds to the addition of an ${\cal
N}=2$ supersymmetric hypermultiplet in the fundamental
representation of the $SU(N_c)$ gauge group.

\begin{figure}[h]
\includegraphics[scale=0.3]{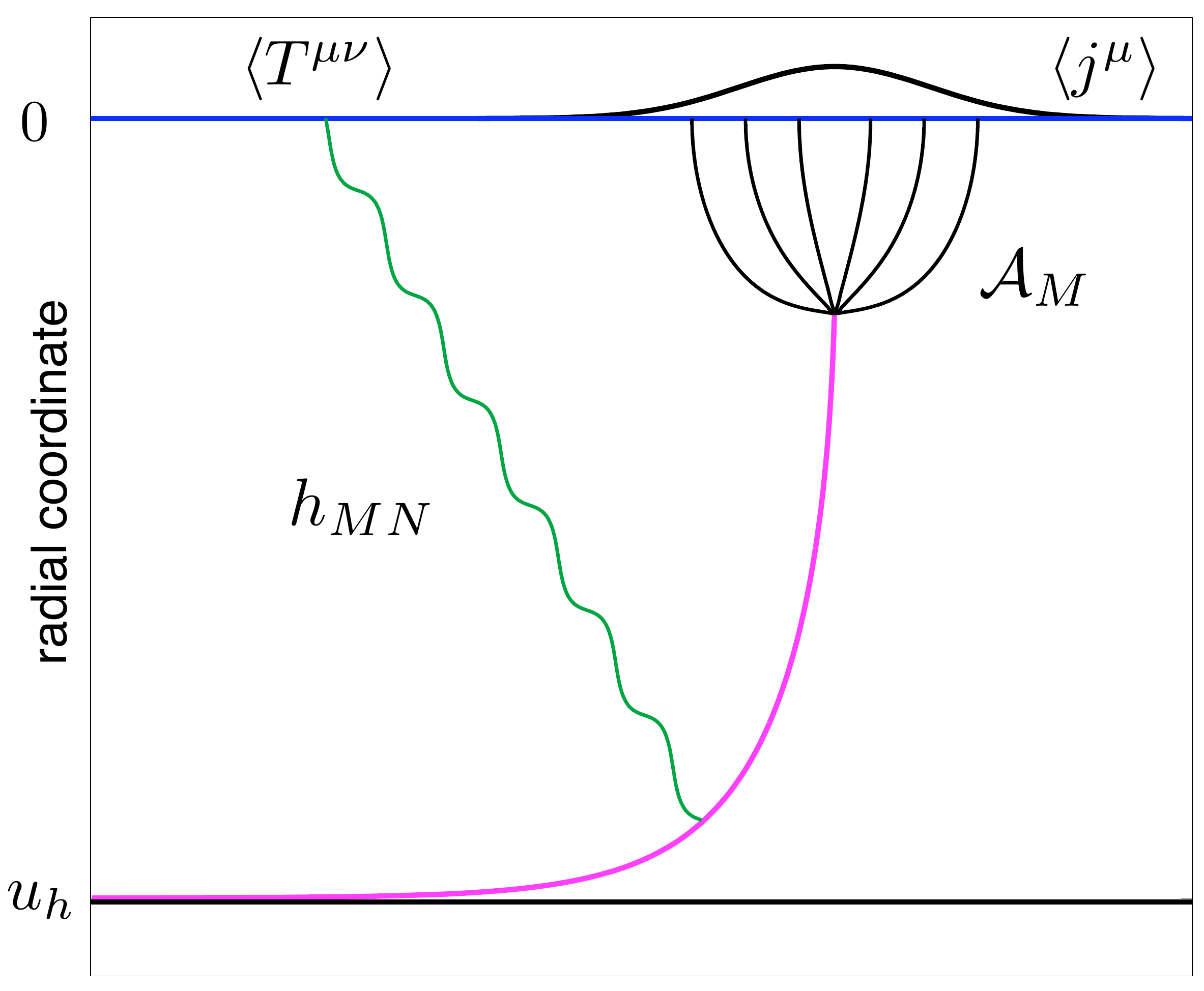}
\caption{\label{bulk2boundary}
    A cartoon of the bulk to boundary problem at finite temperature.
    The endpoints
    of strings are charged under a $U(1)$ gauge field $\mathcal A_M$.
    The boundary of the geometry, located at radial coordinate $u = 0$, behaves like a
    perfect conductor and consequently, the presence of the string endpoints
    induces a mirror current density $j^\mu$ on the boundary.  Via the AdS/CFT correspondence,
    the induced current density has the interpretation of the baryon current of a jet.
    Similarly, the presence of the string induces a perturbation $h_{MN}$ in the metric of the
    bulk geometry.  The behavior of the metric perturbation near the boundary encodes the
    information contained in the perturbation in the SYM stress tensor due to the
    presence of the jet.
    }
\end{figure}

In the large $N_c$ limit the coupling $e^2$ between the gauge field
$\mathcal A_M$ and the string scales as $e^2 \sim 1/\Nc$
and consequently the backreaction of the gauge
field on the string's motion can be neglected.  Similarly, in the
large $\Nc$ limit the $5d$ gravitational constant $\kappa_5^2 \sim
1/N_c^2$ so the backreaction of gravity on the string can be
neglected too.  In short, in the large $\Nc$ limit the string falls
freely under the influence of gravity and the behavior of the gauge
field $\mathcal A_M$ (and the perturbation to the metric $h_{MN}$
caused by the string) can be computed perturbatively. The dual field
theory lives on the boundary of the asymptotically AdS geometry. The
boundary behaves like an ideal electromagnetic conductor, so the
presence of the gauge field $\mathcal A_{M}$ induces a current
$j^\mu(x)$ which resides purely on the boundary.  This is depicted
schematically in Fig.~\ref{bulk2boundary}. Via the AdS/CFT
correspondence, this induced current has the physical interpretation
of the baryon current density of a quark-antiquark pair.  Similarly, the
boundary behavior of the metric perturbation induces a perturbation
in the boundary stress-energy tensor $T^{\mu \nu}$.  The perturbation in
$T^{\mu\nu}$ contains both the stress-energy of the quarks and the
stress-energy of any emitted radiation.

At zero temperature we show that the string endpoint trajectories
generically approach light-like geodesics at late times. While the baryon
density at any point on the boundary may be sensitive to the
behaviour of the source before it relaxes to geodesic motion, we are able to find
boundary observables that are insensitive to early time dynamics.  In
particular, we find that the the center of charge and the 
angular distribution of baryon number
are uniquely determined in
terms of a single parameter $V_s$, which is
the asymptotic spatial velocity of the string endpoint.   The 
angular distribution of baryon number 
falls off as $\frac{1}{(1- V_s \cos \theta)^2}$ with the angle $\theta$ measured
relative to the main axis of the jet.  

At finite temperature we consider string configurations
where the string endpoints travel very far in the spatial directions.
To generate such configurations, the strings must start 
very close to the boundary of the AdS-BH geometry and
must have a very high energy and momentum.  For such
string configurations, we find that the endpoint motion
is approximately geodesic even at early times.  The
corresponding baryon density can be made highly
localized for a period of time 
and hence can be considered to be the baryon density
of a quasi particle.  We track the evolution
of the baryon density from the initial event where
a quark-antiquark pair was created to thermalization of the resulting
jets and the onset of hydrodynamics.

An outline of our paper is as follows.  In Section
\ref{conventions} we write down the conventions
we use.  In Section \ref{fallingstrings} we discuss
the dynamics of falling strings at both zero temperature
and finite temperature.  In both cases approximate
analytic solutions to the string equations of motion
are constructed and in both cases we perform
numerical studies of the string solutions.
In Section \ref{EM} we study the $5d$ electromagnetic
problem and compute the baryon density induced in the
boundary field theory.
In Section
\ref{discussion} we discuss the implications of our results
and in Section \ref{conclusions} we conclude.

\subsection{Conventions}
\label{conventions}

Five dimensional AdS coordinates will be denoted by $X_M$ while
four dimensional Minkowski coordinates will be denoted  by $x_\mu$.
Upper case Latin indices $M,
N, P, \dots$ run over $5d$ AdS coordinates,
while Greek indices run over the $4d$
Minkowski space coordinates.  Worldsheet coordinates will be denoted as
$\sigma^a$ with $a = 0,1$.  The time like world sheet coordinate is $\tau \equiv \sigma^0$
while the spatial coordinate is $\sigma \equiv \sigma^1$.  The endpoints of the string
are located at $\sigma = \sigma^*$.
We choose coordinates such that the metric of the AdS-Schwarzschild
(AdS-BH) geometry is
\begin{equation}
\label{metric}
    ds^2 = \frac{L^2}{u^2}
    \left [-f(u) \, dt^2 + d \x^2 + \frac{du^2}{f(u)} \right ] ,
\end{equation}
where $f(u) \equiv 1-(u/u_h)^4$ and $L$ is the AdS curvature radius.
The coordinate $u$ is an inverse radial coordinate;
the boundary of the AdS-BH spacetime is at $u = 0$
and the event horizon is located at $u=u_h$,
with $T \equiv (\pi u_h)^{-1}$ the temperature of the SYM plasma.

\section{Falling Strings}
\label{fallingstrings}

The dynamics of a classical string are governed by the Nambu-Goto
action
\begin{equation}
\label{ng}
S_{\rm NG} = -T_0 \int d\tau d\sigma \sqrt{-\gamma}
\end{equation}
where $T_0 = \sqrt{\lambda}/2 \pi L^2$ is the string tension, $\sigma$ and $\tau$ are
worldsheet coordinates and $\gamma = \det \gamma_{ab}$ with $\gamma_{ab}$ the
induced worldsheet metric.  The string profile is determined by a set of
embedding functions $X^M(\tau,\sigma)$.   In terms of these functions
we have
\begin{equation}
\gamma_{ab} = \partial_a X \cdot \partial_b X,
\end{equation}
and
\begin{equation}
-\gamma = (\dot X \cdot X')^2 - \dot X^2 X'^2
\end{equation}
where $\dot X^M \equiv \partial_\tau X^M$ and $X'^M \equiv \partial_\sigma X^M$.

The equations of motion for the embedding functions and the open
string boundary conditions follow
from setting the variation of the Nambu-Goto action to vanish.  We write
the equations of motion below in Sections \ref{zeroTasm} and \ref{finiteTasm}.
The open string boundary conditions
require the string endpoints to move at the local speed of
light and that their motion is transverse to the string.

We consider initial conditions such that at worldsheet time $\tau = 0$, the
string is mapped into a single point in space.  We furthermore consider
string profiles which lie in the Minkowski spatial direction $x \equiv x_{||}$.
The point-like initial conditions are
\begin{align}
t(0,\sigma) = 0,
\ \
x(0,\sigma) = 0,
\ \
u(0,\sigma) = u_0.
\end{align}
The remaining necessary initial data is that of $\dot t$, $\dot x$ and
$\dot u$ at $\tau = 0$.  However one of these three
functions may be eliminated via gauge fixing.  For example one
may choose to work in the gauge $\tau = t.$
We leave the detailed discussion of our particular choice
of the remaining initial data for Section \ref{numerics}.  We simply note here
that we can choose initial conditions such that (i)  at zero temperature the string
endpoints move infinitely far apart as $t \rightarrow \infty$ and, (ii) at
finite temperature the initial conditions can be chosen such that the endpoints
travel an arbitrarily large (but finite) distance in the Minkowski spatial directions.
As discussed in Section \ref{EM}, at zero temperature strings whose endpoints
separate to spatial infinity correspond to jets in the dual field theory.
At finite temperature,  strings whose endpoints can be made to travel arbitrarily far
in the Minkowski spatial directions correspond to quasi-particle states in the dual
field theory.

\subsection{Approximate Solutions}
\label{approxSolns}
At both zero temperature and finite temperature it is
possible to obtain approximate solutions to the string
equations of motion.  Our zero temperature solution
is valid in the $t \rightarrow \infty$ limit.  Such a limit
is meaningful to consider at zero temperature as the string
solutions will expand indefinitely and will correspond
to jets in the dual field theory which separate forever.
At finite temperature, the $t \rightarrow \infty$ limit
of string solutions is not a useful limit to consider.  This is
because at finite temperature strings will fall asymptotically
close to the black hole as time progresses.  This corresponds
to the thermalization of jets in the dual field theory.  At finite temperature
we perform an expansion which in essence corresponds to studying an
arbitrarily high energy string.

Both zero temperature and finite temperature approximations
rely on the fact that there exist exact solutions to the string
equations of motion where the string is everywhere moving at the
speed of light. Such solutions have infinite energy and hence are unphysical by themselves, but nevertheless
serve as good solutions to base an asymptotic expansion around.
In our numeric studies shown below in Section \ref{numerics}, we demonstrate
excellent agreement between our asymptotic solutions and the exact numerical
solutions for a wide class of initial conditions.

\subsubsection{Zero Temperature}
\label{zeroTasm}

To obtain the $T = 0$ late time asymptotic solutions, we choose worldsheet
coordinates such that $\tau = t$ and $\sigma = \varphi$
where $\varphi$ is an angular coordinate defined by
\begin{eqnarray}
x &=& \R(t,\varphi) \cos\varphi
\\
u &=& \R(t,\varphi) \sin \varphi.
\end{eqnarray}
The domain in the $(t,\varphi)$ plane in which $ \R(t,\varphi)$
is defined is bounded by two curves $\varphi_s(t)$, $s = 1,2$.
These curves correspond to the motion of the string endpoints and
are determined by imposing the open string boundary conditions.
The open string boundary
conditions require the endpoints to move at the speed of
light and for the endpoint velocity to be transverse to the string.  This implies
that along the curves $\varphi_s(t)$
\begin{eqnarray}
\label{lightlike}
G_{MN}\frac{dX^M}{dt} \frac{dX^N}{dt} &=& 0,
\\ \label{transverse}
G_{MN}\frac{d X^M}{dt}\frac{\partial X^N}{\partial \varphi} &=& 0
\end{eqnarray}

The equation of motion for $\R(t,\varphi)$ is
\begin{align}
\nonumber
\R^2(\R^2 + \R'^2) \ddot \R &- \R^2 (1-\dot \R^2)\R''
+2 \R^2 \R' \cot \varphi + 2 \R'^3 \cot \varphi
\\ \label{reqm}
&-2 \R^2 \dot \R \R' (\dot \R \cot \varphi +\dot \R')
-\R^3 (1-\dot \R^2) = 0 \, .
 \end{align}
Generic open string solutions to the above equation
will fall forever into the bulk of the AdS geometry and will approach
the speed of light.  In particular,
$\R$ will generically scale like $t$ at late times.
Because we are interested in string solutions which correspond to jets
in the dual field theory, we seek solutions whose endpoint separation in the spatial directions
also grows like $t$ in the large $t$ limit.
Such string solutions will expand
at a rate approaching the speed of light, and correspondingly, as time
progresses neighboring points on string will become causally disconnected.
Just as in inflation, any short wavelength structure
in the string will expand and become
long wavelength at late times and correspondingly, $\R'/\R \rightarrow 0$ as $t \rightarrow \infty$.

Our starting point for the $T = 0$ late time asymptotics
is to expand $\R(t,\varphi)$ in an asymptotic expansion about $t = \infty$.
We write
\begin{equation}
\label{Rexpansion}
\R(\varphi,t) =  \Ro(\varphi)  t + \sum_{n = 0}^{\infty} \R_{n}(\varphi) t^{-n}.
\end{equation}
Using the assumption that $\R'/\R \rightarrow 0$ as $t \rightarrow \infty$, one can
conclude that $\Ro(\varphi) = $ constant.  However for completeness we choose to
demonstrate that this is the unique solution to the equations of motion
which satisfies the open string boundary conditions.

Consistency of the above asymptotic expansion (\ref{Rexpansion}) with
the boundary conditions in Eqs.~(\ref{lightlike})--(\ref{transverse})
implies that the functions $\varphi_s(t)$ must have the asymptotic expansions%
\begin{equation}
\varphi_s(t) = \sum_{n=0}^\infty \varphi^{s}_{ n} t^{-n}.
\end{equation}
Note that this expansion does not contain any positive powers of $t$.  Any positive
powers of $t$ in the expansion of $\varphi_s(t)$
would imply superluminal motion for the endpoints.

Substituting the expansion (\ref{Rexpansion}) into the equation of motion
(\ref{reqm}) and setting the leading coefficient in the large $t$ expansion to
vanish, we find the following equation for $\Ro(\varphi)$
\begin{align}
\label{R0}
- \Ro (1-\Ro^2) \Ro'' \,+ \,& 2 (\Ro')^3 \cot \varphi
-2 \Ro^3 (\Ro')^2
\\ \nonumber
-\,&(1-\Ro^2) \left (1-2 \Ro^2 \Ro' \cot \varphi \right ) = 0.
\end{align}
Similarly, substituting the expansions into the boundary conditions
(\ref{lightlike}) and (\ref{transverse}) and setting the leading coefficient in
the large $t$ expansion to vanish, we find the following equations
\begin{eqnarray}
\Ro(\varphi^s_0) &=&1,
\\
\Ro'(\varphi^s_0) &=&0.
\end{eqnarray}
As advertised, the solution to Eq.~(\ref{R0}) which satisfies the above boundary conditions
is simply
\begin{equation}
\Ro(\varphi) = 1.
\end{equation}
We note that the leading order solution $\R = t$ represents a null string with $\gamma = 0$.
This solution simply represents an arc which uniformly expands at the speed of light.

At next to leading order in the $1/t$ expansion, we find the following equation
of motion
\begin{align}
\R_0 \R_1 +\R_2 + 2 \R'_0 \R_1 \cot \varphi + (\R'_0)^3 \cot \varphi
+\R'_0 \R'_1 - \R_1 \R''_0 = 0.
\end{align}
The solution to the above equation simply relates
$\R_2$ to $\R_0$ and $\R_1$ via an algebraic equation.
This feature also occurs at higher orders in the $1/t$ expansion --- all
$\R_n$ with $n>1$ are determined by $\R_0$ and $\R_1$ and their derivatives.
At leading order, the boundary
conditions (\ref{lightlike}) and (\ref{transverse}) yield the equations
\begin{eqnarray}
\varphi^s_1 &=& \R'_0(\varphi^s_0),
\\ \label{rbc}
\R_1(\varphi^s_0) &=&- \frac{1}{2}\left  (\R'_0(\varphi^s_0) \right )^2 .
\end{eqnarray}

The fact that the asymptotic solutions contain two essentially
arbitrary (up to boundary conditions) functions $\R_0(\varphi)$
and $\R_1(\varphi)$ might seem peculiar at first glance.  However
we remind the reader that the initial conditions for a point-like
string configuration contain two function which are also arbitrary up
to boundary conditions.  Evidently these two initial functions must
map onto $\R_0(\varphi)$ and $\R_1(\varphi)$ via the equation
of motion (\ref{reqm}).  Unfortunately we do not currently see
any nice analytic way of relating the initial
data for a point-like string to the functions $\R_0(\varphi)$ and $\R_1(\varphi)$.

{} From the above discussion we see that at late times the strings
endpoints follow the trajectories
\begin{eqnarray}
\label{xgeo}
x_s(t) &=& V_s  t
\\ \label{x5geo}
u_s(t) &=& \sqrt{1-V_s^2} t,
\end{eqnarray}
where $V_s =  \cos \varphi_0^s$.
These trajectories are simply geodesics in the zero temperature AdS geometry.
It is very easy to understand the origin of the asymptotic endpoint geodesic behavior.  
After a long period of time the presence of the gravity
implies that the radial coordinate of the endpoint must increase like $u_s \sim t$. 
However, by assumption are considering strings whose endpoint motion also scales like 
$x_s \sim t$ at late times.  These two conditions together with
the condition that the endpoints move at the 
speed of light then fixes the asymptotic endpoint motion to be a light-like geodesic.

\subsubsection{Finite Temperature}
\label{finiteTasm}

Our approach to obtaining finite temperature
approximate solutions to the string equations of
motion is less systematic than our approach at zero
temperature.  In particular we do not perform a
late time expansion.  The reason is simple --- at
asymptotically late times the string endpoints
always fall into the black hole, and consequently the
endpoint motion in the spatial directions ceases. This
corresponds to the energy loss and thermalization of
jets in the dual field theory.  The late time behavior of any
string solution at finite temperature simply consists of a
string of constant spatial extent, whose radial coordinate
asymptotically approaches the event horizon.

Instead of performing a $t \rightarrow \infty$ expansion, we
construct an asymptotic expansion for a string solution with very
high energy. The validity of our expansion will turn out to be
equivalent to the statement that the energy density and energy flux
on the string are parametrically high.  From the field theory point
of view this is not surprising.  The quark that the string is
describing loses energy to the SYM plasma at a rate $d E_{\rm
quark}/d x \sim E^n$ for some $n>0$. Therefore asymptotically high
energy states will deposit a large amount of energy per unit length
into the plasma.  From the gravitational point of view this
corresponds to having a large energy flux and energy density along
the string.

Generically the endpoint motion of a falling string is bounded
by that of a light-like geodesic which starts at $u = u_0$.  It is easy
to work out the equations of motion for light-like geodesics in the AdS-BH
geometry.  They read
\begin{eqnarray}
\label{dxgeo}
\left (\frac{dx_{\rm geo}}{dt} \right )^2 &=& \frac{f^2}{\r^2},
\\ \label{dugeo}
\left(\frac{du_{\rm geo}}{dt} \right )^2 &=& \frac{f^2 \left(\r^2 - f\right)}{\r^2}
\end{eqnarray}
where $\r$ is a constant which specifies the geodesic.  Note that
for any geodesic which obtains a minimal radial coordinate $u_{\rm min}$
on its trajectory,
the parameter $\r$ must satisfy $\r^2 > f(u_{\rm min})$.
Dividing
Eq.~(\ref{dxgeo}) by Eq.~(\ref{dugeo}) we see
\begin{equation}
\label{dxgeodu}
\left (\frac{dx_{\rm geo}}{du} \right )^2 = \frac{1}{\r^2 - f}.
\end{equation}
Eq.~(\ref{dxgeodu}) may be integrated to determine
the total spatial extent that the geodesic travels.
In the limit $\r^2  \rightarrow f(u_0)$ and $u_0 \rightarrow 0$,
this distance scales like $u_h^2/u_0$.

We consider string configurations where
the spatial component of the string endpoint motion
$x_{s} \approx t$ for times
 $t \ll u_h^2/u_0$.
Under this condition
the endpoint velocity in the Minkowski spatial
directions will be very close to the local speed of light.
Because open string endpoints must always travel at the
speed of light, the velocity in the radial direction must therefore be
small and correspondingly, the radial coordinate of the
string endpoints must change very slowly for
times $t \ll u_h^2/u_0$.

Just as at zero temperature, the assumption that the
$x_{s} \approx t$
implies the string must stretch and expand
as time evolves.  For sufficiently smooth initial conditions%
\footnote
  {
  Initial conditions where there is an arbitrarily
  large amount of structure in the string at early times do not have to relax to
  a smooth string profile after any given amount of time $\Delta t$.
  While the initial structure will be inflated as time progresses,
  because the
  string endpoint can only travel a distance $\sim u_h^2/u_0$, one can
  always cook up initial conditions such that the initial structure never relaxes
  in this time interval.  We do not consider such initial conditions in this
  paper.
  }
this implies short wavelength perturbations in the initial structure
of the string will expand and become long wavelength, resulting in a
smooth string profile at late times. Moreover, as the string
endpoints separate, the middle of the string must fall rapidly
toward the event horizon.  In our numerical studies discussed in
Section \ref {numerics}, we have found that this typically occurs
over a time scale $\Delta t \sim u_h$.  
This is the time scale for a geodesic with $\r \ll 1$ to fall into the 
horizon.  We see this for both of the numerically
generated string solutions in Figs.~\ref{finiteTsymSeq} and \ref{finiteTasymSeq}.
The origin of this behavior can be understood as follows.
Consider the string at times shortly after the creation event.  It will have expanded to
a size $\sim t$.  By construction the string will have a very large momentum density
in the spatial directions.  The momentum density must be very inhomogeneous so that 
each endpoint can move off in its own direction.  
As time progresses the parts of the string with the highest momentum density
will evolve to be located near the string endpoints.  Portions of the string with 
low momentum density will lag behind the endpoints
and fall relatively unimpeded toward the horizon.
After the middle of the string falls into the event
horizon, the trajectories of the two halves of the string which
extend up from the horizon will be uncorrelated. As time progresses,
each half of the string will translate along in the spatial
directions and the endpoint will slowly fall.  We see this nicely for the numerical
strings in Figs.~\ref{finiteTsymSeq} and \ref{finiteTasymSeq}.

With the above motivation in mind,
our starting point for the construction of approximate
finite temperature string solutions is to
consider a half string which asymptotically approaches the event
horizon at spatial distances far from the string endpoint.
We look for solutions which approximately translate at constant
velocity in the spatial directions.  These solutions will be
approximately stationary in the rest frame of
the string endpoint during the time interval $u_h \ll t \ll u_h^2/u_0$.

We choose worldsheet coordinates $\tau = t$, $\sigma = u$.  In these
coordinates the embedding functions are determined by one function $x(t,u)$.
The domain in the $(t,u)$ plane in which $x(t,u)$
is defined is bounded by a curve $u_s(t)$.  This curve corresponds
to the trajectory of the string endpoint and is determined by imposing
the open string boundary conditions.
The equation of motion of $x(t,u)$ is
\begin{align}
\label{xeqm}
2 u \left(1+f x'^2 \right ) \ddot x - 2 u f \left(f - \dot x^2 \right) x''- 4 u f x' \dot x \dot x'  &
\\ \nonumber
+ \,  4 f \left (2-f \left (1-x'^2 \right ) \right ) x'
-4 \left (3-2 f \right) x' \dot x^2 & = 0.
\end{align}

We seek a solution to Eq.~(\ref{xeqm}) of the form
\begin{equation}
\label{perturbativesolution}
x(t,u) = x_{\rm steady}(t,u) + \epsilon x_1(t,u) + \mathcal O(\epsilon^2)
\end{equation}
where
\begin{equation}
\label{sstate}
x_{\rm steady}(t,u) = \rr t + x_0(u),
\end{equation}
is a steady state solution to the equations of motion
with $\rr$ a constant and $\epsilon$
is a formal expansion parameter.
Similarly, we write
\begin{equation}
\label{us}
u_s(t) = u^s_0(t) + \epsilon u^s_1(t) + \mathcal O(\epsilon^2).
\end{equation}
The function
$x_1(t,u)$ characterizes the perturbations
in the string which have inflated to long wavelengths.

We note that having nontrivial time dependence in
$u_s(t)$ does not contradict the steady state assumption
in Eq.~(\ref{sstate}).  As we shall soon see, the steady state
condition implies that $u_0(t)$ must satisfy a geodesic equation.

At leading order in $\epsilon$ the equation of motion (\ref{xeqm})
lead to the following equation of motion for $x_0(u)$
\begin{align}
2 u f \left (\rr^2 - f \right ) x''_0 + \ & 4 f^2 x'^2_{0}
\\ \nonumber
+ \, 4 \big [ (2-f) f -\ & \rr^2 (3 - 2 f) \big ] x'_0 = 0.
\end{align}
The general solution to the above equation is given
by functions which satisfy
\begin{equation}
\label{x0p2}
\left (\frac{\partial  x_0}{\partial u} \right )^2 = \frac{ u^4 \left (\rr^2 - f \right )}{u_h^4 f^2 \left (1- C f \right)}
\end{equation}
where $C$ is an arbitrary constant.
Before integrating
Eq.~(\ref{x0p2}), we first determine $C$ by imposing the
open string boundary conditions.  The reader may note that Eq.~(\ref{x0p2}) is
the equation of motion for the dragging string of Refs~\cite{Herzog:2006gh,Gubser:2006bz}.  There, the string obeys a
Dirichlet condition at the endpoint and consequently $C$ is found to be $1/\r^2$.  We will find a different
value of $C$ because we solve the problem for different boundary conditions.  As a result, we generally
obtain a different solution than the string of Refs~\cite{Herzog:2006gh,Gubser:2006bz}.

With our choice of worldsheet coordinates,
the open string boundary conditions
are simply
\begin{eqnarray}
\label{lightlikeT} G_{MN}\frac{dX^M}{dt} \frac{dX^N}{dt} &=& 0,
\\ \label{transverseT}
G_{MN}\frac{d X^M}{dt}\frac{\partial X^N}{\partial u} &=& 0,
\end{eqnarray}
where all quantities are evaluated along the curve
$u_s(t)$. At leading order in $\epsilon$
these boundary conditions lead to the two equations
\begin{eqnarray}
\label{bconstraint1}
\left (\frac{\partial  x_0}{\partial u} \right )^2 &=& \frac{\rr^2 - f}{f^2},
\\ \label{bconstraint2}
\left (\frac{d u_0^s}{dt} \right )^2 &=& \frac{ f^2 ( \rr^2 - f )}{\r^2},
\end{eqnarray}
where all quantities are evaluated along the curve $u_0^s(t)$.
Comparing Eq.~(\ref{bconstraint1}) with Eq.~(\ref{x0p2}), we
see the two equations agree if $C = 1$.  Furthermore,
comparing Eq.~(\ref{bconstraint2}) with Eq.~(\ref{dugeo}), we see that
to leading order in $\epsilon$, the string endpoints follow light-like geodesics.

With $C = 1$ Eq.~(\ref{x0p2}) may be integrated.  Taking the negative
root of the equation so that the string profile trails the endpoint, we
can find the profile in terms of Appell hypergeometric functions. For the special
case of $\rr \rightarrow 1$, $x_0(u)$ reduces to
\begin{equation}
x_0(u) \stackrel{\rr \rightarrow 1}{=} \frac{u_h}{2}
\left [ \tan^{-1} \frac{u}{u_h} + \frac{1}{2} \log \frac{u_h - u}{u_h + u} \right ],
\end{equation}
which is the well know trailing string profile of Refs \cite{Herzog:2006gh,Gubser:2006bz}.

Just as at zero temperature, the leading order solution to the string
equations of motion is a null string with $\gamma(x_{\rm steady}) = 0$.
Including the $\mathcal O(\epsilon)$
correction to the string profile
implies $\gamma(x(t,u)) = \mathcal O(\epsilon)$.  The energy density
and energy flux of the string both are proportional to $1/\sqrt{-\gamma}$.
Therefore the validity of our asymptotic expansion implies the energy density and
energy flux of the string are very high.

It is straightforward to work out the equations of motion
for the perturbation $x_1(t,u)$.  We do not carry out this exercise
here but we simply note that one can show that the perturbations
do not become large as time progresses.  In particular if $\epsilon$ is small,
then $\epsilon x_1(t,u)$ remains small for all times.  We therefore expect
the perturbative expansion in Eq.~(\ref{perturbativesolution})
to be valid outside the time window
$u_h \ll t \ll u_h^2/u_0$.  Indeed as we discuss in Section \ref{numerics},
the agreement between the steady state result and our
numerical results agrees for all times $t \sim $ a few  $u_h$.

\subsection{Numerical String Solutions}
\label{numerics}

In addition to studying the asymptotic behaviour of falling strings via the
expansions in Section \ref{approxSolns}, we can numerically solve the string equations
of motion to find the worldsheet for arbitrary initial conditions.  These solutions allow us to test the
accuracy of the expansions in Section \ref{approxSolns} and to study string behaviour
in the regimes where the expansions do not hold.

For numerical reasons discussed in Ref \cite{Herzog:2006gh}, we have found it
convenient to work with the Polyakov action instead of the Nambu-Goto
action.  The Nambu-Goto action is classically equivalent to the Polyakov action
\begin{equation}
\label{polyakov}
S_{\rm P}=-\frac{T_0}{2}\int d\tau d\sigma \sqrt{-\eta}\eta^{a b}\partial_{a}X^{M}\partial_{b}X^{N}G_{MN},
\end{equation}
where $\eta_{ab}$
is the worldsheet metric.
The equations of motion follow by varying of the action with respect to $X^M$ and $\eta_{ab}$.
The variation with respect to $X^M$ yields the equation of motion
\begin{equation}
\partial_{a}\left[ \sqrt{-\eta}\eta^{ab}G_{MN}\partial_{b}X^{N}  \right]
=\frac{1}{2}\sqrt{-\eta}\eta^{ab}\frac{\partial G_{NP}}{\partial X^{M}}\partial_{a}X^{N}\partial_{b}X^{P},
\end{equation}
together with the open string boundary condition
\begin{equation}
\label{bc2}
\pi^\sigma_{M}(\tau,\sigma^*) =0,
\end{equation}
where
\begin{equation}
\pi^\sigma_{M}(\tau,\sigma) \equiv \frac{\delta S_{\rm P}}{\delta X'^M(\tau,\sigma)}=
 -T_0 \sqrt{-\eta} \eta^{\sigma a} G_{MN}\partial_a X^N
\end{equation}
and $\sigma = \sigma^*$ at the string endpoints.

The variation of the action with respect to
$\eta_{ab}$ yields the worldsheet constraint
\begin{equation}
\label{constraint}
\gamma_{ab}
=\frac{1}{2}\eta_{ab}\eta^{cd}\gamma_{cd}.
\end{equation}
The constraint equation implies
\begin{equation}
\sqrt{-\gamma} \gamma^{ab} = \sqrt{-\eta} \eta^{ab},
\end{equation}
which when substituted back into the Polyakov action,
easily shows the classical equivalence to the Nambu-Goto
action.

One may fix the worldsheet coordinates $(\tau,\sigma)$ by making a
choice for the worldsheet metric $\eta_{ab}$.  Following Ref
\cite{Herzog:2006gh}, we choose to fix
\begin{equation}
(\eta_{ab})=\left(\begin{array}{cc} -\Sigma(x,u) & 0 \\ 0 & \frac{1}{\Sigma(x,u)}\end{array}\right),
\end{equation}
where $\Sigma(x,u)$ is a stretching function, which we take
to be a function of $x(\tau,\sigma)$ and $u(\tau,\sigma)$ only.
With this choice of worldsheet metric, the worldsheet constraints (\ref{constraint}) read
\begin{align}
\label{con1}
\dot X \cdot X' = 0,
\\ \label{con2}
\dot X^2 + \Sigma^2 X'^2 = 0.
\end{align}
Since we choose to study point-like initial conditions, we must choose initial
data $\dot X^M(0,\sigma)$
consistent with the constraint equation Eq.~(\ref{con2}) and boundary conditions Eq.~(\ref{bc2}).  This
is generically satisfied by specifying $\dot t$ in terms of $\dot x$ and $\dot u$ via
\begin{equation}
f \dot t^2 =\dot x^2 + \frac{\dot u^2}{f},
\end{equation}
in addition to requiring the boundary conditions
\begin{equation}
\dot{t}'(\sigma^*)=\dot{x}'(\sigma^*)=\dot{u}(\sigma^*) = 0
\end{equation}
at $\tau = 0$, where $\sigma^*$ denotes the endpoints of the string.

To choose the functions $\dot x(0,\sigma)$ and $\dot u(0,\sigma)$ as well as
the initial radial coordinate $u_0$, we must consider what kind of states we wish to
study in the field theory.  We first consider the zero temperature theory.
As we discuss below in Section \ref{EM}, the endpoints of the strings roughly
correspond to the locations of the jets in the boundary field theory.  Field theory
states with jets that propagate to infinity therefore correspond to
falling strings whose endpoints escape to infinity in the spatial directions.  We
therefore choose to study string initial conditions
with enough energy and momentum so that the endpoints separate
from each other for all times.
In the finite temperature theory, the distance the endpoints can separate is bounded.
The field theory states here have a natural quasiparticle description until the jets
thermalize.
We therefore choose to study initial conditions where we can tune how far
the jets travel, i.e. how long the quasiparticle lives before thermalizing.

Beyond these restrictions, the initial conditions are not further
constrained by physical requirements.  Different initial conditions
correspond to different structures of the initial string.
What we want to be able to do is to identify aspects of the
resulting jets that are universal, but also aspects that depend
crucially on the choice of initial conditions.

\subsubsection{Summary of Results}
\label{nStringSummary}
We have examined a number of initial conditions in the zero and finite temperature cases and
in each case have confirmed the approximations in Section \ref{approxSolns}.
In particular, the string endpoints asymptote to geodesics in each geometry
and fluctuations along the string at early times are inflated in wavelength
as the string falls.

\begin{figure}[h]
\includegraphics[scale=0.8]{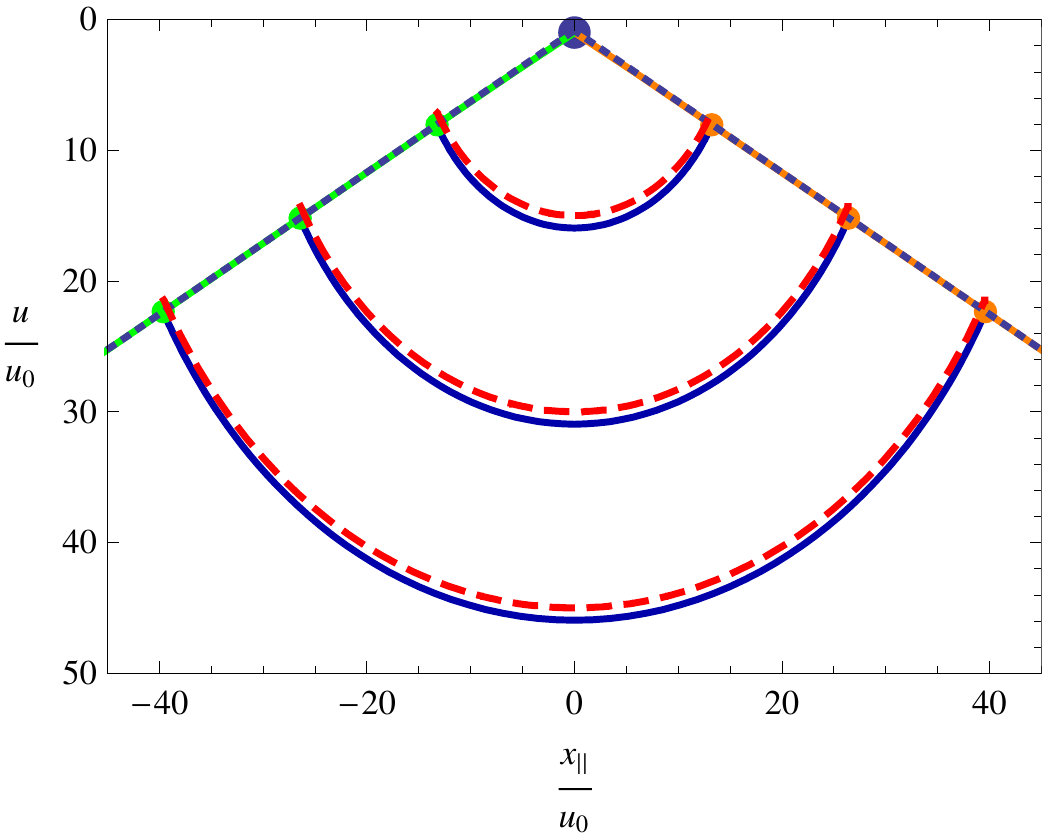}
\caption{
  \label{zeroTsymSeq}
   A falling string at zero temperature.  The solid blue
   curve represents the string at three successive times
   $t_1 = 15u_0, \, t_2 = 30u_0,$ and $t_3 = 45u_0$.
   The dashed red curve represents the null string approximation
   $\R = t$ at the same three times.  The orange and green solid
   curves represent the trajectories of the string endpoints while
   the dotted blue lines represent the geodesic fit the the endpoint
   trajectories.
   The numerical string was
   generated with the initial conditions in Eq.~(\ref{zeroTsymIC})
   at the large blue dot near the top of
   the plot.  As time progresses, the difference between
   the null string and the numeric string becomes small
   compared to the overall size of the string.
}
\end{figure}

We will now showcase four representative solutions --- two at zero temperature and two at finite
temperature.  The details of how these strings are generated
are given below in Section \ref{nStringDetails}.
The first string we examine is plotted in
Fig.~\ref{zeroTsymSeq}, with the null string solution $\R=t$
overlaid.
{} From the figure, we see that the the difference between the null
solution and the numeric solution becomes small compared to the
size of the string as time progresses.  Moreover, we see that the
endpoint trajectory is almost completely geodesic.

\begin{figure}[h]
\includegraphics[scale=0.8]{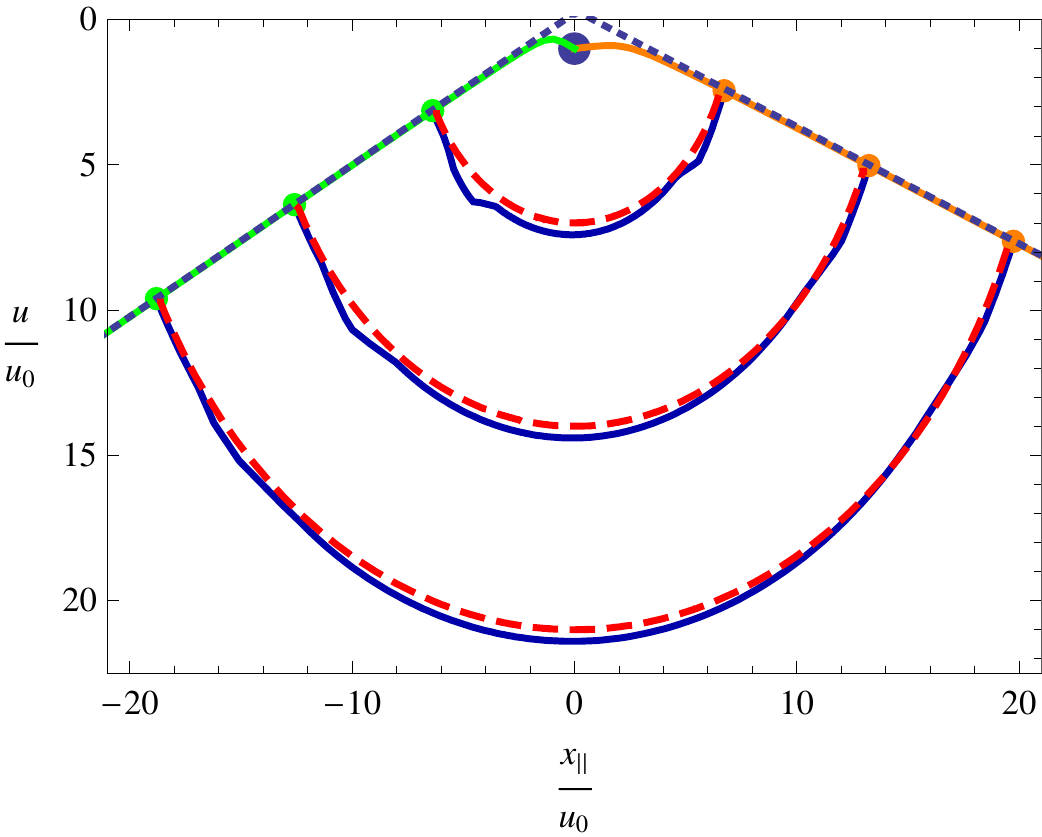}
\caption{
  \label{zeroTasymSeq}
  An asymmetric string at zero temperature generated
  from the initial conditions Eq.~(\ref{zeroTsymIC}).
  The solid blue
   curve represents the string at three successive times
   $t_1 = 7u_0, \, t_2 = 14u_0,$ and $t_3 = 21u_0$.
   The dashed red curve represents the null string approximation
   $\R = t$ at the same three times.  The orange and green solid
   curves represent the trajectories of the string endpoints while
   the dotted blue lines represent the geodesic fit to the endpoint
   trajectories.
   As time progresses,
   the perturbations in the string profile inflate and become long wavelength.
   Correspondingly, the difference between
   the null string and the numeric string becomes small
   compared to the overall size of the string.
}
\end{figure}

The above string was generated with rather symmetric initial conditions.
To better test the series solution Eq.~(\ref{Rexpansion}), we therefore
consider a second zero temperature string with large
fluctuations at early times.
Such a string is shown in Fig.~\ref{zeroTasymSeq}.
\begin{figure}[h]
\includegraphics[scale=0.8]{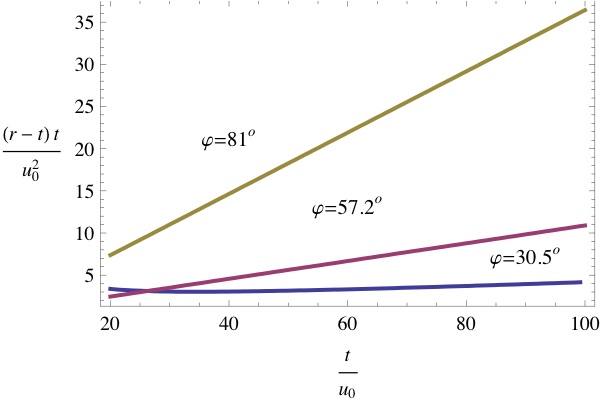}
\caption
  {\label{zeroTasymDelta}
  A plot of the deviation $(\R(t,\varphi)-t) \, t$ of the string shown in Fig.~\ref{zeroTasymSeq} from
  the null string.  According to the asymptotic expansion given
  in Eq.~(\ref{Rexpansion}), for constant values of $\varphi$,
   $(\R(t,\varphi)-t) \, t$ should be a linear function of $t$ at late times.
  The linearity of the plots at late times reinforces the form
  of the expansion in Eq.~(\ref{Rexpansion}).
}
\end{figure}
The profile has two large fluctuations at early times that grow in size
as the string falls.  At these early times, the circular arc solution is not
a good fit to the string profile; there is a large perturbation.
However as time progresses the short wavelength
structure in the perturbation inflates to long wavelengths
and the size of the perturbation becomes small compared to the
growing size of the string.  Again from the figure, one can see
that the endpoints approach light-like geodesic
trajectories.

According to the expansion given in Eq.~(\ref{Rexpansion}), the difference
between any expanding string and the null string $\R=t$ should have the late
time form
\begin{equation}
\R(t,\varphi)-t=\R_0(\varphi)+\frac{\R_1(\varphi)}{t},
\end{equation}
so for fixed $\varphi$, $(\R(t,\varphi)-t) \, t$ should be linear
in $t$ as $t \rightarrow \infty$.  This observation provides a
simple test of the form of the asymptotic expansion given in
Eq.~(\ref{Rexpansion}).
We plot $(\R(t,\varphi)-t) \, t$ in Fig.~\ref{zeroTasymDelta}
for the above asymmetric string configuration shown in Fig.~\ref{zeroTasymSeq}.
As is evident from the figure, for fixed $\varphi$, $(\R(t,\varphi)-t) \, t$
is linear at late times, reinforcing the form of the expansion
given in Eq.~(\ref{Rexpansion}).

\begin{figure*}[htc]
\includegraphics[scale=1.5]{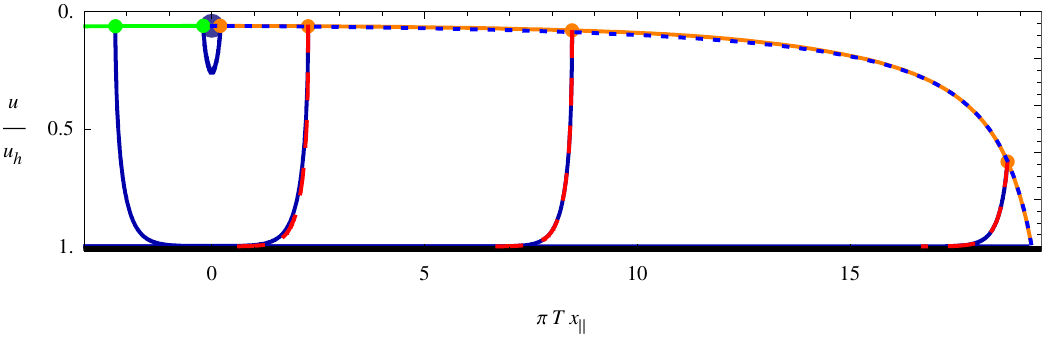}
\caption
  {
  \label{finiteTsymSeq}
  A symmetric string at finite temperature generated from the initial conditions
  Eq.~(\ref{finiteTsymIC}).  The string, shown as the solid blue curve,
  is shown at successive times $t_1=1/5 \pi T,$ $t_2=2.3/\pi T,$ $t_3=8.5/\pi T$, and $t_4=18.8/\pi T$ with the
  corresponding null string, shown as the dashed red curve, plotted at the same times.  The solid
  green and orange curves represent the endpoint trajectories while the dotted blue curve represents
  the geodesic fit to the endpoint trajectory.  The null string agrees very well
  with the numeric string at times $t \sim $ a few $u_h$.  The geodesic fit to the endpoint
  trajectory was obtained by matching to the total distance traveled by the endpoint.
}
\end{figure*}

In the same spirit as with the zero temperature plots above, we exhibit two finite temperature strings --- one smooth
string  solution
and one solution with large fluctuations at early times. A smooth string is shown in Fig.~\ref{finiteTsymSeq}.
It is a symmetric state about $x=0$ whose endpoints travel a spatial distance of $19.25/\pi T$.  The dotted
blue line in the plot represents a geodesic fit to match the stopping distance
of the string endpoint.  The dashed  red curve shows the corresponding null string plotted on top
of the numerical string.  As is evident from the figure, the null string fits the numerical solution very well
for times $t \sim$ a few $u_h$.

\begin{figure*}[htc]
\includegraphics[scale=1.5]{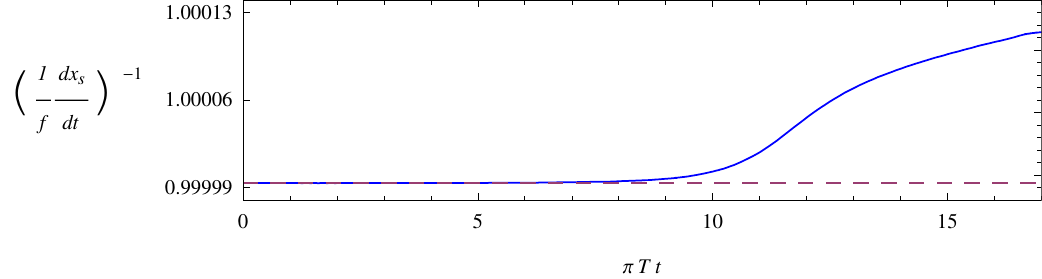}
\caption
  {
  \label{finiteTsquiggle}
  The quantity $f (dx_s/dt)^{-1}$ for the string of Fig.~\ref{finiteTsymSeq}
   is plotted with the solid blue line as a function of time.
   For a geodesic, this is the conserved quantity $\r$.
   We also plot the geodesic fit to the endpoint trajectory with the dashed purple line: $\r=0.999994$.
   As is evident, $f (dx_s/dt)^{-1}$ is approximately constant over the trajectory of the string endpoint.
   The deviations of $f (dx_s/dt)^{-1}$ from a constant give a measure of the expansion parameter
   $\epsilon$ given in Eq.~(\ref{us}).
   }
\end{figure*}

To see how well the endpoint motion is approximated by a geodesic, we plot $f (dx_s/dt)^{-1}$
in Fig.~\ref{finiteTsquiggle}.  For a geodesic, this is the constant $\r$.
The geodesic fit to the stopping distance is plotted in addition to the endpoint data.
In the plot, we see that $f (dx_s/dt)^{-1}$ changes by roughly one part in ten thousand
over the course of the trajectory of the endpoint.  This is completely consistent
with the expansion given in Eq.~(\ref{us}).  In particular, the value of
$f (dx_s/dt)^{-1}$ should change by an amount parameterized by $\epsilon$.
Evidently, in this case $\epsilon \sim 10^{-4}$.

\begin{figure}[h]
\includegraphics[scale=0.8]{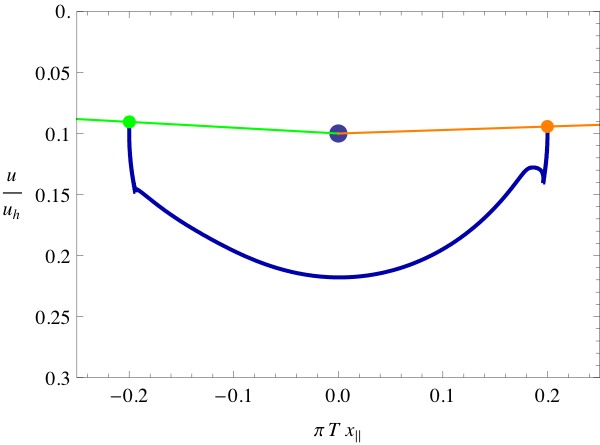}
\caption
  {
  \label{finiteTasymInit}
  A closeup of the string in Fig. \ref{finiteTasymSeq} at time $t=1/5\pi T$.
  This plot shows the initial structure in the string which is inflated to long wavelengths at late times.
}
\end{figure}

\begin{figure*}[htc]
\includegraphics[scale=1.6]{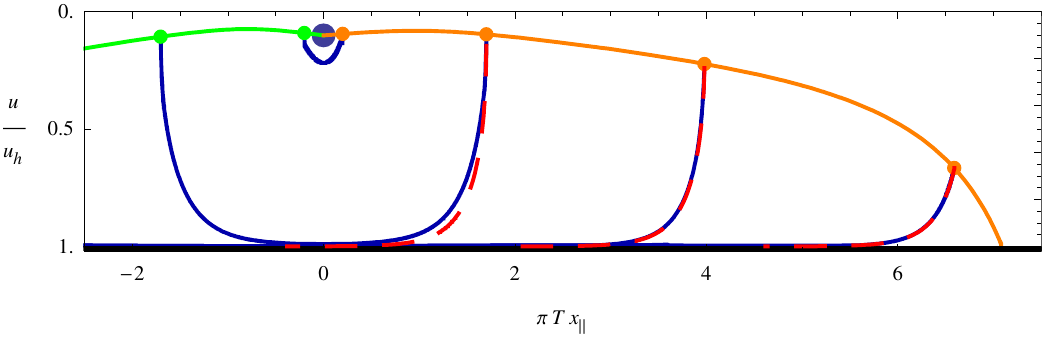}
\caption
  {
  \label{finiteTasymSeq}
  An asymmetric string at finite temperature generated from the initial conditions
  Eq.~\ref{finiteTasymIC}.  The solution and the fitted null string is shown in the same manner as in Fig.~ \ref{finiteTsymSeq}
  at times $t=1/5\pi T, \, 1.7/\pi T,\, 4/\pi T,$ and $6.7/\pi T$.
}
\end{figure*}

We also exhibit an asymmetric finite temperature string, as shown at several
times in Fig.~\ref{finiteTasymSeq}.
A close-up
of the early-time profile is plotted in Fig. \ref{finiteTasymInit}.
As with the asymmetric zero temperature string in Fig.~\ref{zeroTasymSeq}, the early-time
fluctuations quickly inflate to long wavelengths.  The string
quickly relaxes over a time $t\sim u_h$ to match a null string with fluctuations,
matching the predictions of Section \ref{finiteTasm}.

\subsubsection{Numerical Details}
\label{nStringDetails}
We generate our numerical strings by solving the string equations of motion with a canned PDE solver:
Mathematica's \textit{NDSolve}.  We do so with the same technology used in Ref \cite{Herzog:2006gh}.  The
crucial difference with the numerical work in Ref \cite{Herzog:2006gh} is that we must optimize the stretching
function on a case-by-case basis.  We essentially tune the worldsheet metric to each solution so that
gradients along the string are minimized at all times in the string's evolution.  We do so with
a stretching function of the form
\begin{equation}
\label{genericStretch}
\Sigma(x,u)=(x^2+1)^m\left(\frac{1-u}{1-u_0} \right)^n\left(\frac{u_0}{u} \right)^p.
\end{equation}
Our zero temperature strings can generally be tracked with high accuracy if we only turn on the $\frac{u_0}{u}$
part of this stretching function.  Both of the zero temperature strings plotted in this paper simply had $p=2$
and in each case we were able to track the string down to distances past $10^3 u_0$.  The configuration
exhibited in Fig.~\ref{zeroTsymSeq} was generated with the initial conditions
\begin{subequations}
\label{zeroTsymIC}
\begin{eqnarray}
\dot{x}&=&4 u_0 \cos\sigma \exp \left[ -\textstyle{\frac{1}{2}}(1-\cos^2\sigma) \right], \\
\dot{u}&=&2u_0,
\end{eqnarray}
\end{subequations}
where we have specified the endpoint locations $\sigma^*$ to be $0$ and $\pi$ for all of our numerical strings.

Here, we would like to give the reader some intuition about
why we chose these initial conditions by highlighting their salient features.
First, the exponential factor in $\dot{x}$ allows us to control
how much momentum and energy is initially located near
the string endpoints.  This should roughly correspond to how much momentum and energy are stored near the
quark and antiquark at early times.  For this solution, the string therefore has more (but not much more) energy
near its endpoints than in the middle.  The rest of the initial conditions are chosen to be as simple as possible
such that the endpoints fly apart from each other at $t=0$.

The asymmetric string shown in Fig.~\ref{zeroTasymSeq} was generated with a more complicated set of initial
conditions given by
\begin{subequations}
\label{zeroTasymIC}
\begin{eqnarray}
\dot{x}&=&-4u_0 \left(\cos\sigma+\textstyle{\frac{1}{3}}\cos 3\sigma\right), \\
\dot{u}&=&u_0\left(\textstyle{\frac{1}{2}} -2\cos 4\sigma-\cos 5\sigma  \right).
\end{eqnarray}
\end{subequations}
Here we chose initial conditions so that the initial velocities along the string, $dx/dt=\dot{x}/\dot{t}$
and $du/dt=\dot{u}/\dot{t}$, would have large fluctuations at early times.  These evolve to the bumps
that inflate in size as the string falls.

Similar to the first zero temperature solution, the initial conditions for the symmetric finite temperature string in Fig. \ref{finiteTsymSeq} were
\begin{subequations}
\label{finiteTsymIC}
\begin{eqnarray}
u_0&=&\textstyle{\frac{3}{50}}u_h, \\
\dot{x}&=&626 u_0 \cos\sigma\exp \left[-16(1-\cos^2\sigma)   \right], \\
\dot{u}&=&u_0\sqrt{f(u_0)}(1-\cos 2\sigma).
\end{eqnarray}
\end{subequations}
In this case, we needed to turn on the other parts of the stretching function Eq.~(\ref{genericStretch}).  For this string, we found
the choice $m=0.6,\, n=1$, and $p=2$ to be effective.  Also, the initial conditions here are somewhat more extreme than in
the zero temperature cases above.  Our choice reflects an attempt to make a long-lived string by (i) creating the string
near the boundary, (ii) putting most of the energy and momentum near the endpoints, and (iii) giving the endpoints zero
initial radial velocity.  This last choice allows the string endpoints to travel along an $\r\approx \sqrt{f(u_0)}$ geodesic
for a significant fraction of their trajectory, whereby they travel much further in the spatial direction than an
$\r=1$ geodesic.  See Section \ref{discussion} for more details on this point.

The initial conditions for the asymmetric finite temperature string in Fig. \ref{finiteTasymSeq} were
\begin{subequations}
\label{finiteTasymIC}
\begin{align}
u_0\,&=\,\textstyle{\frac{1}{10}}u_h, \\
\dot{x}\,&=\, 30u_0 \left(\cos\sigma+\textstyle{\frac{3}{10}}\cos 3\sigma-\textstyle{\frac{1}{25}}\right), \\
\dot{u}\,&=\, u_0\sqrt{f(u_0)}\left( \textstyle{\frac{1}{2}}- 2\cos 4\sigma -\textstyle{\frac{1}{3}} \cos 5\sigma\right).
\end{align}
\end{subequations}
We used a slightly different stretching function for this string with $m=0.7$, $n=1.5$, and $p=2$.  Here, the initial
conditions are similar to those used for the asymmetric zero temperature string of Fig.~\ref{zeroTasymSeq} in that we
generate a string that has large fluctuations in the velocities $dx/dt$ and $du/dt$ at small times.  We also create the
string near the boundary with a large amount of energy near the endpoints so that, like the long-lived finite temperature
string in Fig.~\ref{finiteTasymSeq}, the endpoints will travel some long distance before falling into the horizon.

\section{The Electromagnetic Problem and the Boundary Baryon Density}
\label{EM}

In the large $N_c$ limit the action for the $U(1)$ gauge field living on
the D7 brane reduces to that of ordinary curved space Maxwell equations.
We therefore start with the $N_c \rightarrow \infty$ effective action
\begin{equation}
\label{action}
S = S_{\rm EM}  + S_{\rm int}.
\end{equation}
The electromagnetic action is given by
\begin{equation}
S_{\rm EM} = -\frac{1}{4 \e^2} \int d^5 x \sqrt{-G} \,F_{MN} F^{MN}
\end{equation}
where $F_{MN}$ is the field strength tensor corresponding
to the gauge field $\A_M$ and $\frac{1}{\e^2} = \frac{N_c}{4 \pi^2}$.
The interaction action is given by
\begin{equation}
S_{\rm int} = \int d^5x \sqrt{-G} \,  J \cdot \A .
\end{equation}
where $J^{M}$ is the current corresponding to the string endpoints.

The equations of motion for the gauge field follow from taking the variation of the
action (\ref{action}).  The variation of $S$ is
\begin{align}
\nonumber
\delta S =&  \int d^5 x \sqrt{-G} \,\delta \A_{N} \left ( \frac{1}{\e^2}D_{M} F^{MN} - J^N \right)
\\ \label{deltaS}
 &+ \frac{1}{\e^2} \int_{u = \epsilon} d^4 x \sqrt{-G} \, \delta \A_{N} F^{5N}.
 \end{align}
The first term in Eq.~(\ref{deltaS}) gives the equations of motion
for the gauge field
\begin{equation}
\label{eqm}
\frac{1}{\e^2} D_{M} F^{MN} = J^N,
\end{equation}
while the surface term gives the boundary baryon current
density
\begin{eqnarray}
j^{\mu}(t,\x) &=& \lim_{u \rightarrow 0} \frac{\delta S}{\delta \A_\mu(t,\x,u)},
\\ \label{j}
&=& \frac{1}{\e^2} \lim_{u\rightarrow0} \sqrt{-G(t,\x,u)} F^{5 \mu}(t,\x,u).
\end{eqnarray}
We note that the above expression for the boundary current density can also
be obtained by integrating the equations of motion (\ref{eqm})
over a gaussian pillbox enclosing the boundary.

{} From Eq.~(\ref{j}) we see that the boundary charge density
is determined by the near boundary behavior of the radial
component of the electric field
\begin{equation}
E_5 \equiv F_{05}.
\end{equation}
The equations of motion for $E_5$ can easily be worked out from
Eq.~(\ref{eqm}).  Introducing a spacetime Fourier transform, we find the
following equations of motion for the mode amplitudes $E_5(\omega,\mathbf q,u)$
\begin{equation}
\label{E5}
E_5'' + A_5 E_5' + B_5 E_5 = S_5
\end{equation}
where
\begin{align}
A_5 & \equiv \frac{u f' - f}{u f},
\\
B_5 & \equiv \frac{\omega^2 - f q^2}{f^2} - \frac{u f' - f}{u^2 f},
\\ \label{S5}
S_5 & \equiv \frac{\e^2 L^2 \left(2 J_0-i u \omega  J_5-u J_0'\right)}{u^3 f}.
\end{align}

We may solve Eq.~(\ref{E5}) with a Greens function $G(u,u')$ constructed
out of homogenous solutions
\begin{equation}
\label{bulk2bulk}
G(u,u') = g_>(u_>) g_<(u_<)/W(u')
\end{equation}
where $W(u)$ is the Wronskian of $g_>$ and $g_>$.  The appropriate
homogenous solutions are dictated by boundary conditions.  The differential
operator in (\ref{E5}) has singular points at $u = 0$ and $u = u_h$ with exponents
$1$ and $\pm i \omega \uh/4$.  Regularity at the boundary implies $g_<(u) \sim u$ as
$u \rightarrow 0$ while the requirement that the black hole not radiate
\cite{Son:2002sd} implies
that $g_>(u) \sim (u - \uh)^{-i \omega \uh/4}$ near the horizon.  The overall normalization
of $g_<(u)$ may be fixed by requiring $\lim_{u \rightarrow 0} g_<(u)/u \equiv 1$.
{} From Eq.~(\ref{j}) we therefore find that the boundary charge density is given by%
\footnote
  {We note that $S_5 \propto \frac{\e^2}{L}$ so the baryon density is
  independent of both $e$ and $L$ as it must be.
  }
\begin{equation}
\label{rho}
\rho(\omega, \mathbf q) =
\frac{L}{e^2} \int_0^{\uh} du' \, \G(\omega, \mathbf q,u') S_5(\omega, \mathbf q,u')
\end{equation}
where
\begin{equation}
\label{bulk2boundaryprop}
\G(\omega, \mathbf q,u) \equiv \frac{g_>(\omega, \mathbf q,u)}{W(\omega, \mathbf q,u)}
\end{equation}
is the Fourier space bulk to boundary propagator.

\subsection{Zero Temperature Baryon Density}

At $T = 0$ the AdS radial coordinate ranges over the
interval $(0,\infty)$ and $f \equiv 1$.
In this case homogenous solutions to Eq.~(\ref{E5})
may be obtained analytically.  The boundary condition on $g_>(u)$
at $u = \infty$ is $\lim_{u \rightarrow \infty} g_>(u) = 0$.  We find
\begin{eqnarray}
g_<(u) &=& u I_0(u \sqrt{q^2 - \omega^2}), \\
\label{g>}
g_>(u) &=& u K_0(u \sqrt{q^2 - \omega^2}),
\end{eqnarray}
where $I_0(z)$ and $K_0(z)$ are modified Bessel functions.
We note that as given in Eq.~(\ref{g>}),
$g_{>}(u)$ has an ill defined $u \rightarrow \infty$ limit.
This can be remedied by giving $\omega$ an infinitesimal imaginary part.
Causality dictates that the retarded Greens functions be analytic in the upper half
$\omega$ plane, so the appropriate prescription is to send
$\omega \rightarrow \omega + i \epsilon$.  With this prescription $g_{>}(u)$
vanishes at $u = \infty$ for all $q$ and $\omega$.
The resulting bulk to boundary
propagator is therefore
\begin{equation}
\label{b2b0}
\G(\omega,\mathbf q,u) = - K_0(u \sqrt{q^2- \omega^2 }).
\end{equation}
Using the aforementioned $i \epsilon$ prescription, Eq.~(\ref{b2b0})
may be Fourier transformed to real space to obtain
\begin{equation}
\label{G0}
\mathcal{G}(t,\mathbf x,u)=\frac{1}{\pi}\theta(t) \frac{d}{d \w}\delta(\w),
\end{equation}
where $\w$ is defined to be
\begin{equation}
\w = -t^2 + \mathbf x^2 + u^2.
\end{equation}

The zero temperature real space source for $E_5$ is
\begin{equation}
\label{S50}
S_5 = e^2 L^2 \left [ -\partial_u  (J_0/u^2)
+  \partial_0 J_5/u^2 \right ].
\end{equation}
Combing the real space version of Eq.~(\ref{rho}) as well as Eq.~(\ref{G0}) and (\ref{S50})
and integrating by parts, we obtain
\begin{align}
\nonumber
\rho(t,\x) = L^3 & \int du' \, d^4 x'
\Big [\partial_{u'} \mathcal G(t{-}t',\x {-} \mathbf x',u') J_0(t',\mathbf x',u')
\\ \label{rho0}
&+ \ \partial_0 \mathcal  G(t{-}t',\x {-} \mathbf x',u') J_5(t',\mathbf x',u') \Big ] \frac{1}{u'^2}.
\end{align}
Using coordinate time as a parameter for the string endpoint
trajectories, the current corresponding to the string endpoints is
\begin{equation}
 \label{current}
J^M(t,\mathbf x,u) = \theta(t) \sum_s (-1)^s
\frac{1}{\sqrt{-G}} \frac{d X_s^M}{dt} \delta^3(\mathbf x - \mathbf x_s) \delta (u -u_s)
\end{equation}
where the endpoint trajectory is given by $X_s^M(t)$ and the factor of $(-1)^s$
comes from the fact that the string endpoints are oppositely charged.
Substituting Eq.~(\ref{current}) into Eq.~(\ref{rho0})
and integrating over $\mathbf x'$ and $u'$, we obtain
\begin{equation}
\label{rho0_2}
\rho(t,\x) = \sum_s (-1)^s \int dt' \frac{d^2}{d \w_s^2} \delta(\w_s)  \Gamma_s(t,t').
\end{equation}
where
\begin{equation}
\label{Gamma}
\Gamma_s(t,t') = -\frac{2}{\pi} \theta(t-t') \theta(t') u_s(t') \left[ u_s(t') +(t - t')   \frac{du_s}{dt'} \right ].
\end{equation}
and $\w_s$ is evaluated at
\begin{equation}
\w_s = -(t-t')^2 + (\x- \mathbf x_s(t'))^2 + (u_s(t'))^2.
\end{equation}

\subsubsection{Late Time Asymptotics}

In general the value of the baryon density at a particular
point $(t,\mathbf x)$ depends on the history of the sources
in the bulk of the AdS geometry.  However,
the dependence on the history of sources can
be reduced if one considers quantities which are averaged in some sense,
such as moments of the baryon density or the angular distribution
of baryon number discussed below.
Moreover, the fact that the trajectories of
the string endpoints asymptotically approach light-like geodesics
suggests that the late time behavior of some averaged quantities
should be independent of the history of the the bulk sources.  We explicitly
demonstrate that this is true for for the first moment of the baryon density
and for the angular distribution of baryon number defined below in Eq.~(\ref{Bjet}).

Consider the first moment
of the baryon density associated with jet $s$, the baryonic center of charge position
\begin{equation}
\bar {\x}_s(t) \equiv
\frac{ \int d^3 x \, \x \, \rho_s(t,\x) }{ \int d^3 x \, \rho_s(t,\x)}
\end{equation}
where $\rho_s(t,\x)$ is the baryon density corresponding
to a single string endpoint labeled by $s$.  The denominator
is simply $ \int d^3 x \, \rho_s(t,\x) = (-1)^{s+1}$.
This follows from the fact that the mirror charge induced
on the boundary is always opposite of the corresponding source charge
in the bulk.
{} From Eq.~(\ref{rho0_2}) we therefore
have
\begin{equation}
\bar {\x}_s(t) =  -\int dt' d^3 x \, \x \, \frac{d^2}{d \w_s^2} \delta(\w_s)  \Gamma_s(t,t').
\end{equation}
Shifting integration variables and performing a couple of integrations by parts,
we obtain
\begin{equation}
\label{avgx}
\bar {\x}_s(t) = \pi   \int dt' \frac{\partial}{\partial t'}
\left [ \frac{  \mathbf x_s(t')  \Gamma_s(t,t')}{\dot Z_s(t,t')} \right ]
\frac{\theta \big ( Z_s(t,t') \big )}{\sqrt{Z_s(t,t')}}
\end{equation}
where we have defined
\begin{equation}
Z_s \equiv (t-t')^2 - \left (u_s(t') \right )^2
\end{equation}
and $\dot Z_s \equiv \partial Z_s / \partial t'$.

In the limit $t \rightarrow \infty$ the domain of integration in
Eq.~(\ref{avgx}) will be dominated by $t' \sim t$.
As discussed in Section \ref{zeroTasm}
the late time behavior of the string
endpoint motion asymptotically approaches light-like geodesics.  The particular
trajectories are given in Eqs.~(\ref{xgeo})--(\ref{x5geo}).  When
substituted into Eq.~(\ref{avgx}), these trajectories yield the late time
behavior
\begin{equation}
\label{avgx2}
\bar {\x}_s(t) = \mathbf v_s t
\end{equation}
where
\begin{equation}
\label{vsasrel}
v_s = \frac{1}{V_s} + \left (1-\frac{1}{V_s^2}
\right ) \tanh^{-1} V_s,
\end{equation}
and $V_s $ is the asymptotic velocity of the string
endpoint in the Minkowski spatial directions.  (The directional dependence of
$\mathbf v_s$ is the same as that of $\mathbf V_s$.)
We therefore see that  the velocity of the center of
baryon density does not generically equal the velocity of the
the string endpoint in the spatial directions.
However in the $V_s \rightarrow 1$ limit the two velocities do agree.
In particular
\begin{equation}
\label{vgoestoone}
v_s = 1 +\left ( 1 - \log \frac{2}{1-V_s} \right ) (1-V_s) + \mathcal O((1-V_s)^2).
\end{equation}
In this limit the string endpoint trajectory approaches a line of constant radial
coordinate $u$ and the position of the center of baryon density coincides with the
spatial coordinate of the string endpoints.

At zero temperature it is useful to consider the
angular distribution of baryon number
\begin{equation}
\label{Bjet}
\frac{dB_s}{d \Omega} \equiv  \int_0^{\infty} r^2 dr \rho_s(t,\x)
\end{equation}
where $r \equiv | \x|$. This quantity is in spirit similar to the jet energy profiles 
used in collider physics, 
see e.g. \cite{Ellis:1993ik,Seymour:1997kj}.
For localized lumps of baryon density which escape to spatial 
infinity, this function measures the angular distribution of baryon number
of the jet.%
\footnote
  {
  Alternatively, one could measure the angular distribution
  of baryon number at infinity with the integrated baryon
  number flux $r^2 \int dt \, \hat x \cdot \mathbf j(t,\mathbf x)$.  
  }
 
{} From
Eq.~(\ref{Bjet}) and Eq.~(\ref{rho0_2}) we have
\begin{equation}
\label{db}
\frac{dB_s}{d \Omega} = (-1)^s  \int_0^{\infty} r^2 dr dt'
\frac{d^2}{d \w_s^2} \delta(\w_s) \Gamma_s(t,t').
\end{equation}
Using the delta function to carry our the $r$ integral, we obtain%
\footnote
  {Technically we should be summing 
  over several zeros of the delta function in 
  Eq.~(\ref{db}).  However if one assumes
  $\cos \theta < 0$, then there is only
  one zero of the delta function with $r > 0$.
  We consider $\cos \theta < 0$
  and obtain the general result for $\frac{dB_s}{d \Omega}$
  by analytic continuation.
  }
\begin{align}
\label{dB1}
\frac{dB_s}{d \Omega} = 
(-1)^{s+1}     \int dt' 
\frac{  \Delta_s(t,t')-3 x_s(t')^2 \cos^2 \theta }{8  \Delta_s(t,t')^{\frac{5}{2}} } \Phi_s(t,t')
\end{align}
where $\theta$ is the angle between $\mathbf x_s$ and $\mathbf x$ and 
\begin{equation}
\Delta_s(t,t') \equiv  (t-t')^2-  (1-\cos^2 \theta) \, x_s(t')^2 - u_s(t')^2 
\end{equation}
and
\begin{equation}
\Phi_s(t,t') \equiv \Gamma_s(t,t') \theta \big ( (t-t')^2 -x_s(t')^2 - u_s(t')^2 \big ).
\end{equation}

In the $t \rightarrow \infty$ limit the integration in Eq.~(\ref{dB1}) will be
dominated by times $t' \sim t$.  Just as for the 
center of charge, we may therefore use the 
asymptotic endpoint trajectories in Eq.~(\ref{dB1}).
Substituting the trajectories in 
Eqs.~(\ref{xgeo})--(\ref{x5geo})
into Eq.~(\ref{dB1}), we find
\begin{align}
\label{dB2}
\frac{dB_s}{d \Omega} =   (-1)^{s{+}1} \frac{1{-}V_s^2}{4 \pi}
\int_0^{\frac{t}{2}} dt'
\frac{ t t'  (2 t t' {+}2 t'^2 V_s^2 \cos^2 \theta {-} t^2)}
{ \left[ t^2 - 2 t t' + t'^2  V_s^2 \cos^2 \theta  \right ]^{\frac{5}{2}} }.
\end{align}
Carrying out the integration in Eq.~(\ref{dB2}), we find
the remarkably simple result
\begin{equation}
\label{dB3}
\frac{dB_s}{d \Omega} =   (-1)^{s+1} \frac{1-V_s^2}{4 \pi \left (1-V_s \cos \theta \right )^2}.
\end{equation}

{} From Eq.~(\ref{dB3}) we may compute the average opening
angle of the baryon density.  A simple calculation shows that for
each jet the opening angle is given by
\begin{equation}
\bar \theta_s = \frac{\pi}{2 V_s} \left ( \sqrt{1-V_s^2} -1+V_s\right ).
\end{equation}
In the $V_s \rightarrow 1$ limit this becomes
\begin{equation}
\label{exptheta}
\bar \theta_s = \frac{\pi \sqrt{1-V_s}}{\sqrt{2}} +\mathcal O(1-V_s).
\end{equation}
which shows that jets can be made highly collimated and focused.
Moreover, from Eq.~(\ref{vgoestoone}), we see the limit where
jet are focussed coincides with the limit where
the center of baryon density velocity is close to the speed of light.

\subsection{Finite Temperature Baryon Density}

At finite temperature we obtain the boundary
baryon density by numerically solving the Maxwell
equations (\ref{E5}).  For a given momentum $(\omega, \mathbf q)$,
the source $S_5$ defined in Eq.~(\ref{S5}),
is computed by using the
electromagnetic currents corresponding to the numerical
string solutions discussed in Section \ref{numerics}.  In addition, for each
$(\omega, \mathbf q)$ the homogeneous solution
$g_{>}(u)$ is evaluated by numerically integrating
the homogeneous differential equation
(\ref{E5})
(without sources)
outward from the horizon.
Then the solution $g_{<}(u)$ is evaluated by numerically integrating
the same equations (without sources)
inward from the boundary.
Given these numerically determined homogeneous solutions,
the baryon density is evaluated by numerically performing the
radial integrals in Eq.~(\ref{rho}), with the source $S_5$.  The subsequent
Fourier space baryon density is then numerically Fourier transformed to
real space.  In what follows, we consider the string
generated by the initial conditions given in Eq.~(\ref{finiteTsymIC}).

\begin{figure*}[htc]
\includegraphics[scale=0.37]{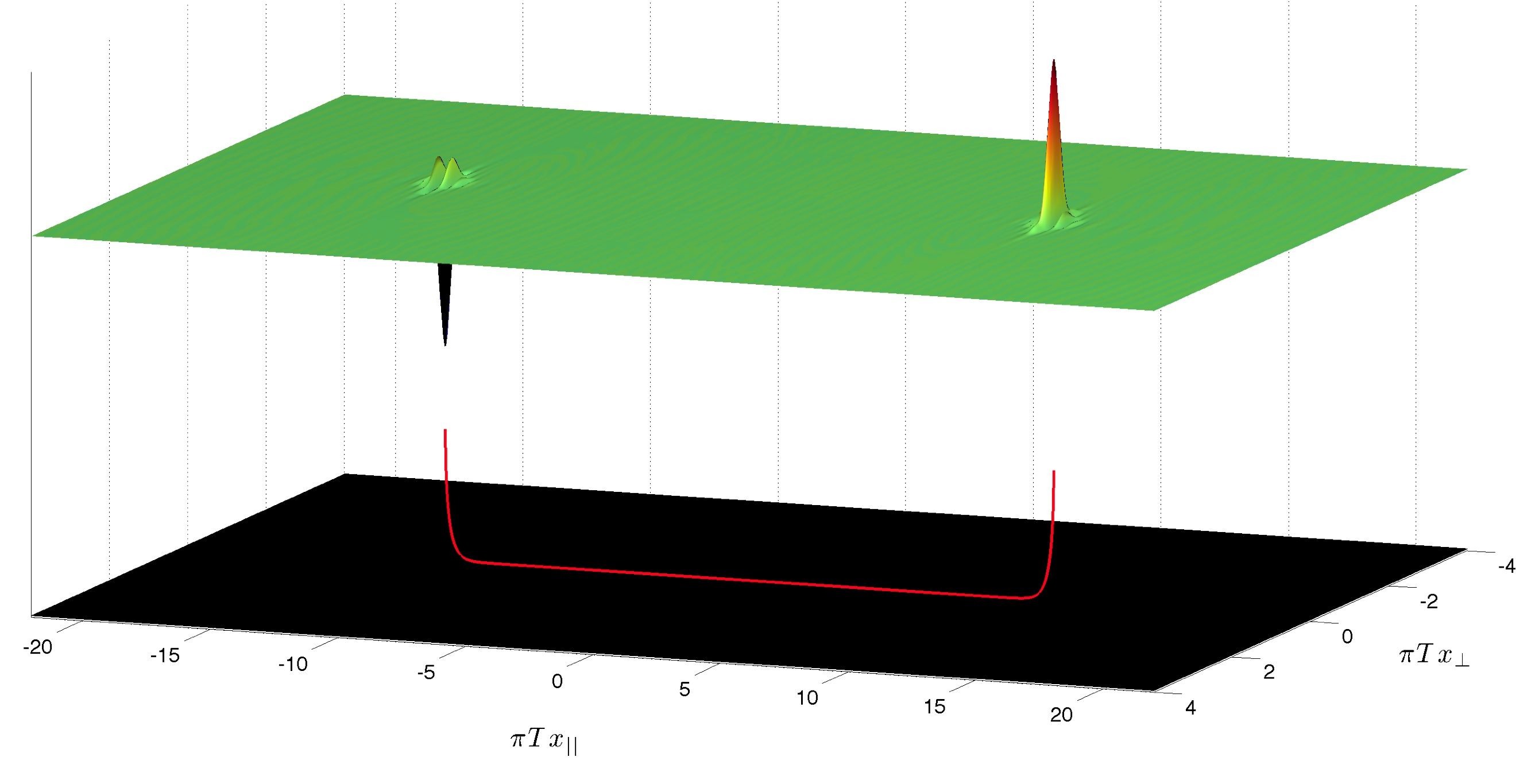}
\caption{\label{stringjet}
    A snapshot of the string profile and the corresponding
    baryon density at an instant of time well before thermalization.
    The black plane represents the event horizon of black hole
    and the surface on the top represents the
    boundary baryon density.  The red curve is the string and
    the vertical axis in the plot serves as both the AdS radial coordinate and the
    magnitude of the baryon density.  The baryon density is
    highly localized above the string endpoints.
    }
\end{figure*}
Fig.~\ref{stringjet} shows a snapshot of a
typical finite temperature string/baryon density configuration
at a time before thermalization, but much after the initial event where the string was
created.%
\footnote
  {See \url{http://www.phys.washington.edu/~karch/Papers/Falling/3D.gif}
  to download a movie of this event.
  }
The black plane at the bottom of the figure
represents the event horizon of the black hole and the surface on the top of the figure represents the
boundary baryon density.  The red curve is the string and
the vertical axis in the plot serves as both the AdS radial coordinate and the
magnitude of the baryon density.  As is apparent from the figure, the baryon
density is highly localized above the string endpoints.  This type of behavior
is universal at times $t$ such that $u_h \ll t \ll u_h^2/u_0$ where
$u_0$ is the initial radial coordinate where the string was created.

The fact that the baryon density is localized above the
string endpoints is easy to understand from the zero
temperature analysis in the previous section.
First of all, in the limit where the sources in the bulk are
very close to the boundary, the electromagnetic bulk to
boundary propagator may be approximated with its zero temperature
limit.  During the time interval $u_h \ll t \ll u_h^2/u_0$, the radial
coordinate of the string endpoint will have not changed
much from its initial value of $u_0$.
In particular, as discussed in Section \ref{finiteTasm}, the trajectories of the
string endpoints are approximately given by light-like geodesics in the
AdS-BH geometry, so from Eq.~(\ref{dxgeodu}), the rate that the endpoint
falls toward the black hole will be suppressed by some power of
$u_0/u_h$.%
\footnote
  {
  For example, for the special case of $\r = 1$, Eq.~(\ref{dxgeodu})
  implies that
  $\dot u_s(t) \approx  \frac{u^2_0}{u^2_h}$
  in the window $u_h \ll t \ll u_h^2/u_0$.
  }
The source $S_5$ for the radial component
of the electric electric field $E_5$ will contain temperature
dependence --- at leading order in $u_0/u_h$ this temperature
dependence is simply the temperature dependence
of the trajectories.  Therefore, to leading order in
$u_0/u_h$ and during the window $u_h \ll t \ll u_h^2/u_0$,
we may use the zero temperature expression Eq.~(\ref{avgx2})
for the center of the baryon density but with the finite temperature
trajectories for the string endpoints.  For string
configurations whose endpoints
propagate very far in the spatial directions, to leading order
in $u_0/u_h$ we have $x_s(t) = t$ during the window
$u_h \ll t \ll u_h^2/u_0$.  From Eq.~(\ref{avgx2}) we therefore conclude
that $\bar { x}_s(t) = t$ as well, showing that
the position of the string endpoints agrees with the center of baryon density.

\begin{figure}[h]
\includegraphics[scale=0.19]{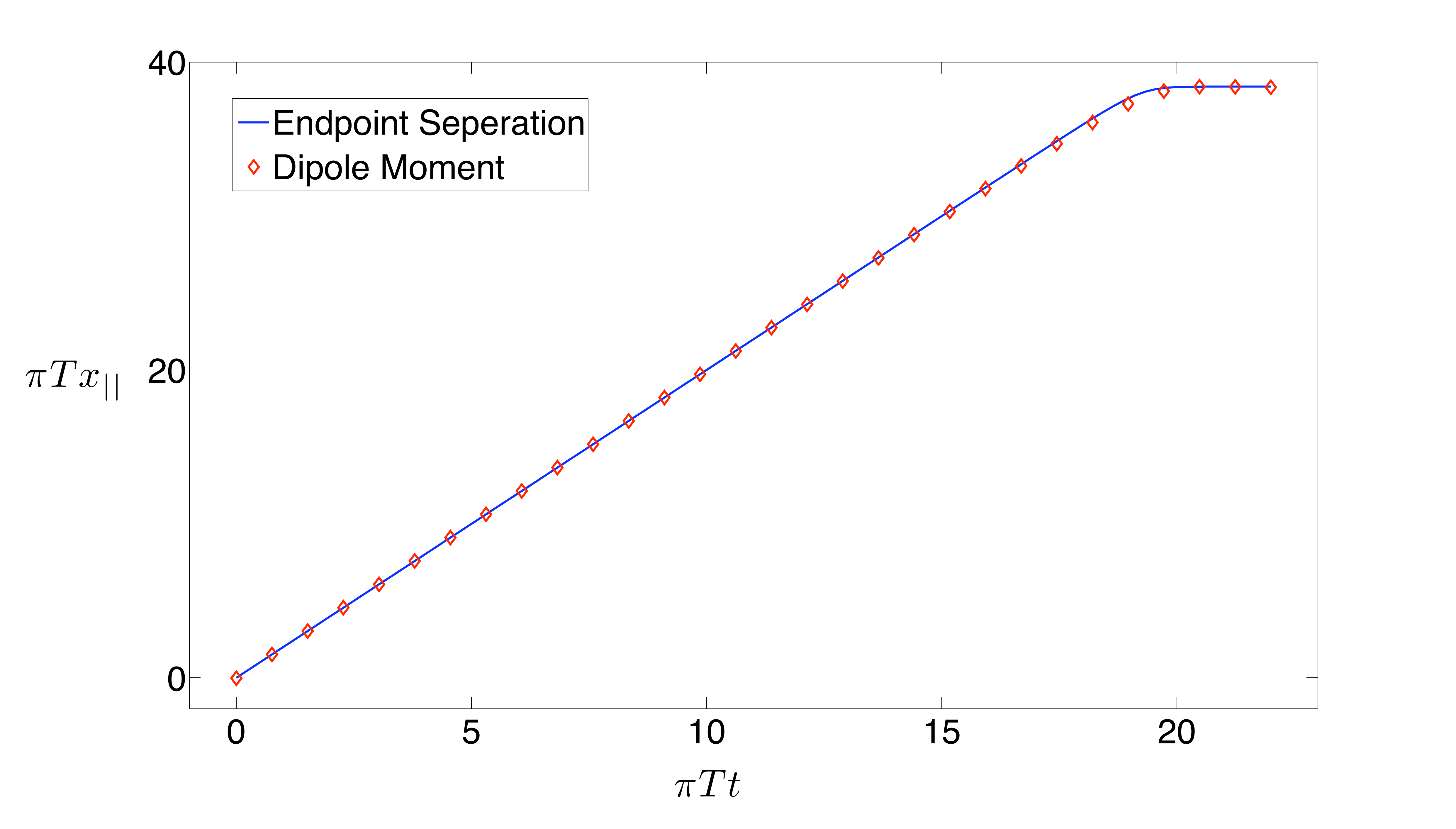}
\caption{\label{dipole}
    A plot of the dipole moment of the baryon number and the string
    endpoint separation in the Minkowski spatial directions.  The
    endpoint separation agrees very well with the dipole moment.
    There is however a small discrepancy near $t \sim u_h^2/u_0$,
    which in this plot is $\pi T t \sim 17$.  This is due to the fact that
    the endpoint velocity slows down at a rate $\sim e^{-4 \pi T t}$
    whereas the velocity of the center of baryon density
    slows down at a rate $\sim e^{-2 \pi T t}$.
    }
\end{figure}

To elaborate on this behavior, we plot in Fig.~\ref{dipole} the
dipole moment of the baryon density and the difference in spatial coordinates
of the string endpoints.  As is evident from the figure, the two curves agree very well
over the vast majority of the string endpoint trajectories.
However while the figure demonstrates that as $t \rightarrow \infty$
the dipole moment agrees with the endpoint separation,
at times $t \sim u_h^2/u_0$, there is a small discrepancy
between dipole moment and the endpoint separation.
As we discuss below, the rate that the dipole moment relaxes to a
constant is a measure of how the baryon density itself relaxes to
its corresponding hydrodynamic modes.  This rate will turn out
to be much slower than the rate that the endpoints slow down,
and this explains the discrepancy in the plot at times $t \sim u_h^2/u_0$.

\begin{figure*}[htp]
\includegraphics[scale=0.35]{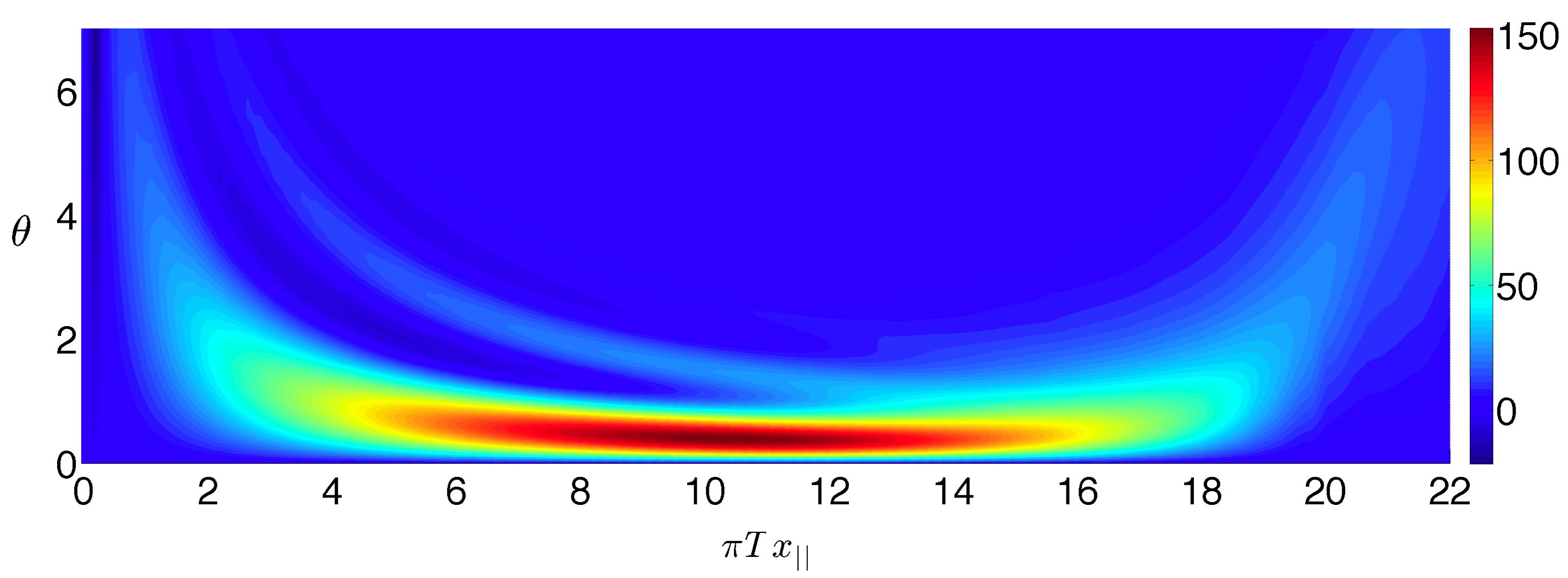}
\caption{\label{dBdt}
    A plot of $dB_s/d \theta$ for a typical string configuration
    where the endpoints travel far in the spatial directions.  At
    times $u_h\ll t \ll u_h^2/u_0$, $dB_s/d \theta$ is highly
    localized about $\theta = 0$.  During these times
    the baryon density wavepacket describes a quasi-particle.
    Around times $t \sim u_h^2/u_0$, which for this example
    is $\pi T t \sim 17$, the quasi-particle has lost
    most of its energy and the resulting jet thermalizes.  Correspondingly,
    the baryon density of the jet spreads out and diffuses at late times.
    Note that the total baryon number contained in any slice of
    constant time is unity.
    }
\end{figure*}

Fig.~\ref{dBdt} shows a plot of the angular distribution of
baryon number
\begin{equation}
\frac{dB_s}{d \theta} \equiv 2 \pi \sin \theta \int_{0}^{\infty} r^2 dr \rho_s(t,\x).
\end{equation}
for a single jet
as a function of both $t$ and opening angle $\theta$.  
At distances close to the initial creation
event $dB_s/d \theta$ is spread out.  This just corresponds to the fact the
lump of baryon density has a finite width and is very close to the origin.%
\footnote
  {There is a potential numerical effect that might be influencing
  the spread of the baryon density at early times.  When solving the
  equation of motion (\ref{E5}) for the radial component of the electric field
  $E_5(\omega,\mathbf q,u)$, we use a finite Fourier transform range with a cutoff $\Lambda$.
  Correspondingly
  in position space the numerically generated
  baryon density cannot have a width that is parametrically
  smaller than $1/\Lambda$.  This effect can potentially make the
  early time behavior of  $dB_s/d \theta$ artificially more spread
  out than it actually is.  We have however checked that
  the early time behavior of $dB_s/d \theta$ remains
  as shown in Fig.~\ref{dBdt} as $\Lambda$ is increased.
  }
In particular, it takes a time $t \sim u_0$ for light to reach the 
boundary after the initial event where the string is created.
After a time $t \sim $ a few $u_0$, distinct wave packets
of baryon density corresponding to each quark will have formed
and each packet will have a width $\sim u_0$.   From Fig.~\ref{dBdt},
we see that after a time $t \sim$ a few $u_h$, $dB_s/d \theta$
has settled down to a relatively constant form.  As discussed in Section
\ref{finiteTasm}, the time scale
$t \sim u_h$ corresponds to the amount of time it takes for the string
to evolve from the point-like initial conditions to the quasi-steady
state configuration given in Eq.~(\ref{perturbativesolution}).
During times in the window $u_h \ll t \ll u_h^2/u_0$,
the baryon density slowly spreads out in the transverse
directions.  In Fig.~\ref{dBdt}, this manifests itself
insofar that the angular width of  $dB_s/d \theta$
is approximately constant.   Again, in this window of time
the temperature dependence in the baryon density
will come from that of the trajectories of the
string endpoints.  That is, to leading order in $u_0/u_h$
the angular width can be computed using the zero
temperature expression for the angular distribution of baryon number in
Eq.~(\ref{dB1}), but with the temperature
dependent endpoint trajectories.   The precise
result for the opening angle is rather complicated and depends on 
on details of the endpoint trajectory (i.e. the exact value of $\r$) so we do not
give it here.

At late times, the baryon density spreads out rapidly.
In the gravitational theory this corresponds to times in which
the string endpoints asymptotically approach the event horizon.
As a string endpoint nears the horizon,
to an observer on the boundary, the electric fields sourced by the string endpoint
will have arbitrarily long wavelengths due to the strong gravitational redshift caused
by the black hole.  Correspondingly the baryon density induced on the boundary
will be diffuse and spread out.  From hydrodynamic considerations, one
expects that the late time evolution of the baryon density will be that of the diffusion equation
\begin{equation}
\label{diffusion}
\left(\partial_0 -D \del^2 \right ) \rho = \Theta
\end{equation}
where $D = 1/2 \pi T$ is the baryon density diffusion constant \cite{Kovtun:2003wp,Myers:2007we} and
$\Theta$ is a localized source in space and time.  Using techniques developed
in Refs \cite{Chesler:2007an,Chesler:2007sv,Gubser:2008vz},
it is easy to compute the source from the gravitational theory.
We have carried out this exercise, and for the symmetric
strings discussed in Section \ref{numerics}, found the following simple result
\begin{equation}
\label{Theta}
\Theta(t,\mathbf x) = \mathbf d \cdot \del \delta^{3}(\mathbf x) \delta (t)
\end{equation}
where $\mathbf d$ is the dipole moment of the baryon density which as mentioned
above, corresponds to the $t \rightarrow \infty$ limit of the string endpoint separation
in the spatial direction.  The above result for the source is easy to understand.
At leading order in gradients,
its form is fixed by the requirement that it be local in space and time
and that the dipole moment of the baryon density in the hydrodynamic regime
simply be $\mathbf d$.

We now turn to the rate in which dipole moment shown in
Fig.~\ref{dipole} relaxes to its final value of $\mathbf d$.  The hydrodynamic
treatment of the baryon density discussed above always
yields a constant dipole moment.  Therefore, at late times, any
time dependence in the dipole moment
is a measure of how quickly the system relaxes to
a hydrodynamic description.
Generically, one expects that the rate that the system relaxes to
hydrodynamics will governed by the
imaginary part of the quasinormal modes in the current-current correlator.
For $\mathcal {N}=4$ SYM, these modes were first found in Ref \cite{Kovtun:2005ev}
and are given by
\begin{equation}
\omega_n=-2\pi T n(\pm 1+i), \label{qnm}
\end{equation}
where $n$ is a positive integer.   We therefore expect the system
to relax to hydrodynamics with a characteristic time scale $1/2\pi T$.

As one can easily
reason from the geodesic equation (\ref{dxgeo}), the rate that the
string endpoints slow down as they approach the event horizon
is
\begin{equation}
\label{stringratex}
\dot x_s(t)  = \frac{f}{\r},
\end{equation}
so $\dot x_s(t) \sim e^{-4 \pi T t}$.
This rate in fact applies under more general circumstances
than for geodesics --- any trajectory which is light-like
will slow down
at this rate near the horizon.
Similarly, the rate that the radial coordinate
increases is
\begin{equation}
\label{stringratex}
\dot u_s(t) \approx f.
\end{equation}
Now suppose the source emits a signal at time $t_0$.  It will take some time
$\tau$ for this signal to reach the boundary.  After a time $\Delta t$ the source will
have fallen a distance $\Delta s \approx \frac{\dot u_s}{f} \Delta t \approx \Delta t$
closer to the horizon.  Therefore
a signal emitted at time $t_0 + \Delta t$ will take a time $\tau + 2 \Delta t$
to reach the boundary.  It therefore follows that an observer on the boundary
will see the string endpoint velocity in the spatial direction
decrease like $e^{-2 \pi T t}$.  We therefore expect that the velocity of the
center of baryon density decrease like $e^{-2 \pi T t}$ instead of
$e^{-4 \pi T t}$.  The time scale
associated with the decay therefore agrees with the lowest quasinormal
mode computed in Ref \cite{Kovtun:2005ev}

The late time behavior of the 
center of charge can also be computed analytically from the bulk
to boundary problem discussed in Section \ref{EM}.
We have carried out this analysis and found that the result
simply reproduces the geometric optics argument given above.

\section{Discussion}
\label{discussion}

A few remarks about our work are in order. We found that universally
the endpoint motion of our strings rapdily settles down to a
lightlike geodesic. Not just that, in fact the whole worldsheet
approaches a lightlike configuration which we were able to construct
analytically. There are two reasons why this was anticipated. For one,
the effective string action goes to zero as the string falls. For
the zero temperature string, the Nambu-Goto action in AdS is
identical to Nambu-Goto in flat space with a position dependent
string tension that scales as $\frac{1}{u^2}$. On top of this, the
effective string tension also decreases as the string becomes more
and more lightlike as it falls. For example, for both for a uniformly translating
string $x=vt$ at zero temperature, and the dragging string of 
\cite{Herzog:2006gh,Gubser:2006bz} at finite temperature, the
on-shell action scales as $\sqrt{1-v^2}$. One can again interpret
this as a velocity dependent string tension that goes to zero as $v$
goes to 1. Indeed, in our late time configuration each point on the
string individually follows a lightlike geodesic. At finite temperature a similar point
was also made in \cite{Gubser:2008as}.

One result we obtained that is perhaps surprising
is that in general, the baryon density on the boundary
is not localized directly above the string endpoints.  In fact
from Eq.~(\ref{vsasrel}) we see that at zero temperature
and late times,
the spatial velocity $V_s$ of the string endpoints
is different than the velocity $v_s$ of the center of baryon
density.  The two velocities
only agrees when $V_s$ approaches the speed of light.
In this limit the baryon density on the boundary is localized
directly above the string endpoints.

\begin{figure}[htp]
\includegraphics[scale=0.18]{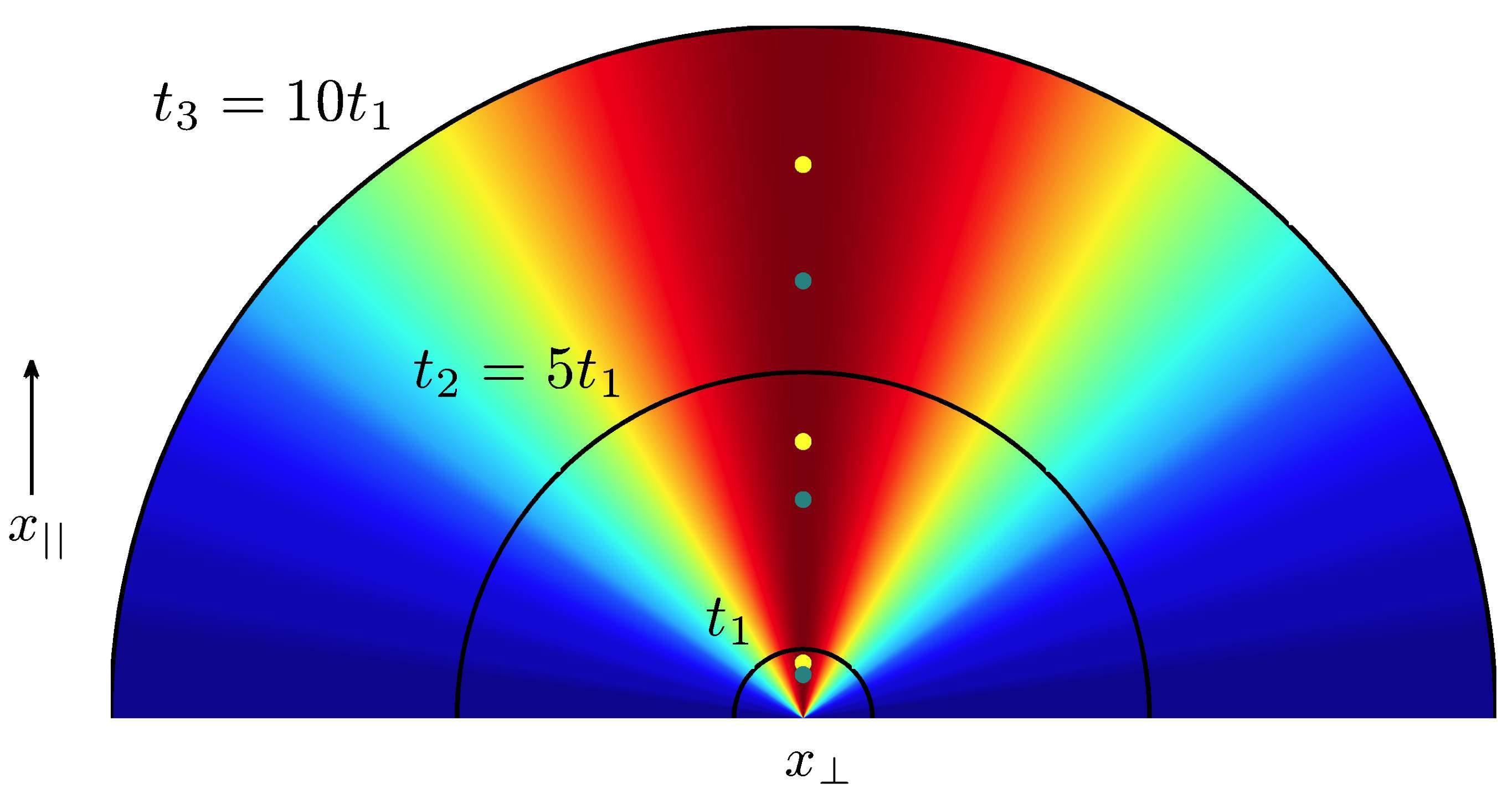}
\caption{\label{geojet}
    A plot of the zero temperature
    baryon density given in Eq.~(\ref{georho})
    at three times $t_1, \ t_2,\ t_3,$ with $V_s = 0.8$.
    At each time, the baryon density is localized on the solid
    black line.  The coloring in the plot shows the angular
    distribution of baryon number, with red showing a high
    concentration and blue showing a low
    concentration.  The yellow dots show the spatial position
    of the charge in the bulk at each time, with the top dot
    corresponding to time $t_3$ and the lowest dot corresponding
    to time $t_1$.  The blue dots
    show the position of the center of baryon density at each time
    with the top dot
    corresponding to time $t_3$ and the lowest dot corresponding
    to time $t_1$.
    As is evident from the plot, the position of the charge in the bulk
    does not coincide with the center of baryon density.  In fact,
    from Eq.~(\ref{vsasrel}) they move at different velocities so
    their separation grows linearly with time.
    }
\end{figure}

To understand the origin of this behavior it is useful to
understand the baryon density induced on the boundary
by charges which exactly follow geodesics
in the zero temperature AdS geometry.  Consider two charges
created at $t = 0$, which then move apart
on the light-like geodesics
\begin{eqnarray}
x_s(t) &=& V_s t,
\\
u_s(t) &=& \sqrt{1-V_s^2} \, t + u_0.
\end{eqnarray}
One can readily substitute the above trajectories
into Eq.~(\ref{rho0_2})
and compute the induced boundary baryon density.
A remarkable feature of the above geodesics is that
the electric fields produced by the
charges at times after the creation event
do not induce any baryon density on the boundary
--- all of the baryon density is induced from the
initial flash of light produced during the creation event.%
\footnote
  {
  More precisely, all of the baryon density comes from
  differentiating the $\theta(t')$ in Eq.~(\ref{Gamma}).
  }
Indeed it has been noted in several papers
\cite{Danielsson:1998wt,Lin:2007fa} that eternal geodesics, or geodesics
which have infinite extent in the past and
future, do not induce boundary densities
at zero temperature.
In the late time limit, the baryon density produced
by the initial flash of light is simply
\begin{equation}
\label{georho}
\rho^{\rm geo}_s(t,\mathbf x)=(-1)^{s+1} \frac{(1-V_s^2)}{4\pi r^2 (1- \hat x \cdot \mathbf V_s)^2}\, \delta(t-r).
\end{equation}
This
equation describes a focused spherical shell of baryon density
expanding at the speed of light.  While the shell of baryon
density itself expands at the speed of light, the center of
baryon density associated with each jet in fact moves at
a speed less than the speed of light.  This is because
the baryon density is expanding in both the transverse and
longitudinal directions.  This is shown in Fig.~\ref{geojet}.
As one can easily compute from
Eq.~(\ref{georho}), the velocity of the center of baryon
density is given by Eq.~(\ref{vsasrel}), and is in general
not equal to $V_s$.

At zero temperature, the fact that
the baryon density induced by
charges moving on geodesics
is completely determined by the initial flash of light
emitted during the creation event might lead
one to suspect that the baryon density
induced by string endpoints, whose trajectories
only asymptotically approach geodesics, is very sensitive to
the initial creation event and not just the asymptotic
endpoint trajectories.  This
intuition is only partially correct.  While the value of the baryon
density at a particular point $(t, \mathbf x)$ might
be sensitive to the details of the creation
event and the histories of sources in the bulk,
certain quantities which are averaged, such as moments
of the baryon density, are not sensitive to the detailed
history of sources in the bulk.  One can explicitly see from
Eq.~(\ref{avgx}) that the late time behavior of the center of baryon density
is completely determined by the late time behavior of the endpoint trajectories.
Furthermore, one can explicitly see from Eq.~(\ref{dB1}) that the late time
angular distribution of baryon number is also completely determined by the asymptotic
behavior of the string endpoints.  Evidently, the information
encoded in the radiation emitted during the times where the string endpoint
trajectories are settling down to geodesics contains
all of the information about the asymptotic behavior of the endpoint motion.

For the case in which one string endpoint approaches a geodesic with $V_s = 0$,
we see from Eq.~(\ref{dB3} that the corresponding angular distribution of baryon
number will be completely isotropic.  Of course for a general value of $V_s$,
one can always boost to a frame where this occurs.
The angular collimation of baryon
density at late times is therefore a consequence of a Lorentz boost.

We note that for the finite temperature string configurations
studied in this paper, the center of baryon density did roughly
coincide with the location of the string endpoint.
We emphasize that the agreement
is largely a consequence of the particular states we chose to study.
We chose to study string configurations where the string endpoints
traveled arbitrarily far in the spatial directions before
falling into the black hole.  Such configurations naturally
have the interpretation of quasi-particles in the boundary field theory.
However, to generate such a configuration, the
endpoint velocity must reside almost entirely in
the spatial directions for most of the trajectory.
Furthermore, the endpoints must move at the
speed of light.  Therefore $V_s \approx 1$ for
most of the trajectory.   The fact that the baryon
density resides over the string endpoint is therefore
completely consistent
with the zero temperature analysis given above.

It would of course be interesting to study the stress-energy tensor
of the jets produced by massless quarks.  The stress-energy tensor of
very massive quarks
moving through a strongly coupled SYM plasma has been studied in Refs
\cite{Chesler:2007an,Chesler:2007sv,Gubser:2007xz,Gubser:2007ga}
while the stress-energy tensor of a meson moving through a
strongly coupled SYM plasma has been studied in Ref \cite{Gubser:2007zr}.
However, while we have found that moments of the baryon density
and the angular distribution of baryon number 
are in general insensitive to the exact initial conditions used to create
the string, we suspect that the structure of the stress-energy
tensor will be very sensitive to the initial conditions.  The reason
for this
is that the strings corresponding to massless quarks in the field
theory approach null strings after the initial
creation event.  As a consequence of this, the determinant
of the worldsheet metric, which vanishes for a null string,
is strongly dependent on the perturbations
which exist on top of the null string. The $5d$ string stress tensor
\begin{equation}
t^{MN}(Y) = -\frac{T_0}{\sqrt{-G}} \int d^2 \sigma
\sqrt{-\gamma}\gamma^{a b}\partial_{a}X^{M}\partial_{b}X^{N}
 \delta^5(Y {-} X),
\end{equation}
will therefore
also depend strongly on the perturbations, and consequently, via
the gravitational bulk to boundary problem, the boundary stress tensor
will as well.  This should be contrasted with the electromagnetic
bulk to boundary
problem analyzed in this paper, where the
electromagnetic currents in the bulk $J^M$ are not
sensitive to the perturbations on top of the null strings.

It would be advantageous to find averaged
quantities related to the boundary stress-energy tensor,
such as moments or jet functions,
which are not sensitive to initial conditions.  If this is
not possible, one can always in principle compute the boundary
stress-energy tensor by in some sense averaging
over a reasonable set of initial conditions or equivalently,
by averaging over
perturbations on top of a null string.  
Whether or
not such an averaging procedure is manageable
remains to be determined.

At finite temperature one quantity which seems to be
somewhat insensitive to the deviations from the null string
is the total stopping distance $\Delta x$ traveled by the quark.
At least at the level of the null string approximation, the stopping
distance is determined by the initial radial coordinate $u_0$
and the geodesic parameter $\r$.  For geodesics which only
fall toward the horizon, the stopping distance can be obtained by integrating
Eq.~(\ref{dxgeodu}),
\begin{equation}
\label{Deltax}
\Delta x = \int_{u_0}^{u_h} du \frac{1}{\sqrt{\r^2 - f(u)}}.
\end{equation}
This quantity is however, very sensitive to the value of $\r$.
For example, consider the case of $\r = 1$.  Then in the $u_0 \rightarrow 0$
limit, Eq.~(\ref{Deltax}) integrates to $\Delta x = \frac{u_h^2}{u_0}$.  Now
consider $\r = \sqrt{f(u_0)}$ and take $u_0 \rightarrow 0$.  This configuration
yields
$\Delta x = \frac{\sqrt{\pi} \Gamma(\frac{5}{4})}{\Gamma(\frac{3}{4})}  \frac{u_h^2}{u_0} = 1.31\frac{u_h^2}{u_0}$.
Therefore a seemingly small change in $\r$ can yield a $30\%$
change in $\Delta x$.  The origin of this behavior is simply that
$\Delta x$ as defined in Eq.~(\ref{Deltax}), is not analytic at
$\r = \sqrt{f(u_0)}$.  Moreover, $\r = \sqrt{f(u_0)}$
sets the endpoint velocity in the spatial direction to be the local speed of light, and
hence can maximize the total distance traveled.

Based on the above arguments, one might suspect that the deviations
from geodesics in the endpoint trajectories
might have an important influence of the total distance
traveled by the endpoint.
{} From Fig.~\ref{finiteTsquiggle} one can see that the quantity, $f (d x_s/dt)^{-1}$,
which is equal to $\r$ for a geodesic, is in fact not constant.
It changes by $1$ part in $10^4$ over the course of the trajectory
of the string endpoint.
This is consistent with the perturbative string solutions
discussed in Section \ref{finiteTasm}.  In general, corrections
to the endpoint trajectories will be $\mathcal O(\epsilon)$.
In fact, the actual value that one chooses for $\r$ is arbitrary up
$\mathcal O(\epsilon)$ corrections.  From Fig.~\ref{finiteTsquiggle}
the possible values of $\r$ range from $\r-1 = 6.5 \times 10^{-6}$
to $\r-1 \approx -10^{-4}$.  For the value of $u_0 = 0.06 u_h$
used in the numerical solution shown in Fig.~\ref{finiteTsquiggle},
these two values for $\r$ lead to values for $\Delta x$ which differ by a factor of
two.  Therefore, if one wants to approximate the endpoint trajectory with
a geodesic, one must carefully choose the value of $\r$.
However, as shown in Fig.~\ref{finiteTsquiggle},
$\r$ is essentially constant for the first half of the trajectory.  It is during
this part of the trajectory that the string endpoint is close to the boundary
and the formula in Eq.~(\ref{Deltax}) is most sensitive to the value of $\r$.
If one uses the initial value of $\r-1 = 6.5 \times 10^{-6}$, then one obtains a distance
traveled by the geodesic which differs from the actual distance traveled by the string
endpoint by $8\%$.

Even if one chooses to use the initial value of $\r$
and approximate the trajectories of the string endpoints
with geodesics,
relating the parameters $\r$ and $u_0$
to the initial quark energy or momentum is in general
nontrivial.  This was discussed in Ref \cite{Gubser:2008as},
where the stopping distance of a gluon was considered.
The analysis of \cite{Gubser:2008as}
tried to relate the parameters specifying
the geodesic to the initial energy of the string.
Without full string solutions only estimates were possible.
Ref \cite{Gubser:2008as}
estimated a stopping distance
$\Delta x = c \, u_h \left ( \frac{2 E_{\rm string}}{T \sqrt{\lambda}} \right )^\frac{1}{3}$
where $c \approx 1$ and $E_{\rm string}$ is the energy
of a string which asymptotically approaches the event horizon.
In order to compare this estimate to our numerical strings,
we take $E_{\rm string}$ to be half of the initial energy
of our point-like strings.  This is reasonable as
the initial point-like configuration rapidly settles down
to a quasi steady state configuration
where the center of the string asymptotically approaches
the event horizon.   For the initial conditions given
in Eq.~(\ref{finiteTsymIC}), this yields $E_{\rm string} \approx 1167 \sqrt{\lambda}T$.
Using the estimate of Ref \cite{Gubser:2008as}, we therefore
obtain a stopping distance of $\Delta x \approx 13.3 u_h$.
The actual distance
traveled by the string endpoint is $\Delta x \approx 19.3 u_h$.
We therefore violate the estimate of Ref \cite{Gubser:2008as}
by $30\%$.  Consequently, we suspect that the analysis of Ref \cite{Gubser:2008as}
is incomplete.

Last, we discuss how our results generalize to string configurations
which lie in multiple spatial directions.   Such string configurations 
give rise to jets which are in general not back to back.  Moreover,
such configurations can give rise to additional jets
caused by energetic gluons.  As discussed in Ref
\cite{Gubser:2008as}, an energetic gluon is represented in gauge/string duality
by a kink in a string.   It is very easy
to extend our analysis to general string configurations.
Consider first the case of finite temperature.  
Suppose we consider a configuration which has in addition to the
two jets caused by the string endpoints, $N$ additional gluon 
jets.  Such configurations will have a very inhomogeneous distribution
of momentum density in the initial string configuration.  Portions of the 
string with a small momentum density will quickly fall to the horizon
over a time scale $t \sim u_h$.  
After this has happened, the trajectories of each portion of string 
extending up from the horizon are uncorrelated.  
The string endpoints and the gluon kinks 
will then move in arbitrary directions and each portion of the string 
profile will be very well approximated by steady state string profile. 
Therefore, the full string profile will look like $2 + N$ trailing strings
which are translating in different directions and are connected
in a region of characteristic size $\sim u_h$.  For times $t \gg u_h$,
the baryon density induced by each string endpoint will therefore
be virtually identical to that obtained by considering strings which
only exist in one spatial direction.

To generalize our zero temperature results, we consider strings 
whose endpoint position in the spatial directions 
grows with time like $|\mathbf x_s | \sim t$.  
Such strings will expand as they fall and 
will also approach a null string configuration.  
Moreover, the radial coordinate of each endpoint will 
increase with time like $u_s \sim t$.  We can therefore repeat 
the late-time expansion of Section \ref{zeroTasm} for this general case.  
Doing so, we find that the light-like boundary condition on each string 
endpoint forces its trajectory to asymptote to a light-like geodesic.  
Correspondingly, the angular distribution of baryon number
for each jet will again be given by Eq.~(\ref{dB1}).  However, 
there will in general be no correlation between the direction of
each string endpoint and the resulting cones of baryon density will thus
point in arbitrary directions relative to each other.  Moreover, the 
string asymptotes to a profile given by an 
arbitrary curve on the 4-sphere $\mathbf x^2 + u^2 = t^2$.  So as in 
the finite temperature case, a general falling string at zero 
temperature may have a number of kinks dual to gluon jets.  

\section{Conclusions}
\label{conclusions}

We have shown, via the AdS/CFT correspondence, that strings
falling in anti de-Sitter space are dual to jets in large $N_c$,
strongly coupled $\mc{N}=4$ super Yang-Mills theory at zero
and finite temperature.  Ultimately, we studied the baryon
density of these jets.  To do this, we obtained the endpoint
motion for the dual strings.  At zero temperature,
a falling string's endpoint always travels a path that asymptotes
to a lightlike geodesic.  The finite temperature case is more
complicated.  Here we find that, for strings whose endpoints travel
far in the Minkowski spatial directions, the endpoint motion approximates a
lightlike geodesic even at early times.  We then numerically confirmed both of
these results.

Next, we computed properties of the baryon density for these jets.
Our results are universal at zero temperature.  In particular,
two integrals of the baryon density --- the center of charge of each jet
and the angular distribution of baryon number  ---
are completely determined by the endpoints' asymptotic trajectories. 
From this, we found that zero temperature jets travel forever and
that their baryon density is focused into two
cones whose angular dependence is related to the 
asymptotic motion of the string endpoints 
via the simple formula Eq.~(\ref{dB3}). 

At finite temperature, we see three time scales in the evolution of the 
baryon density.  The first scale is the initial radial coordinate $u_0$
where the string was created.  This scale sets the duration of time
it takes light to propagate from the bulk of AdS to the boundary
and sets the formation time for the production of quasiparticles.
The second scale is the horizon radius $u_h$.  It takes a time
$t \sim u_h$ for the string in the bulk to relax to a quasi steady
state configuration.  The third time scale is $u_h^2/u_0$. 
For times in the window
$u_h \ll t \ll u_h^2/u_0$, the string uniformly translates along
in the spatial directions
and the endpoint slowly falls toward the black hole --- the radial
coordinate does not change much from its initial value of $u_0$.  
Correspondingly, 
during these times, the baryon
density is highly localized above the string endpoint and 
is naturally interpreted as the baryon density of a quasi particle.
At times $t \sim u_h^2/u_0$, the string endpoint is no 
longer close to the boundary and the resulting baryon 
density starts to spread out.  This corresponds to 
the thermalization of the jet.  For times $t \gg u_h^2/u_0$ the
string endpoint has fallen asymptotically close to the event
horizon and the baryon density evolves according to hydrodynamics
--- it simply diffuses.

\begin{acknowledgments}
We thank S.D.~Ellis, C.P.~Herzog, M.J.~Strassler, and L.G.~Yaffe, for many useful discussions.
This work was supported in part by the U.S. Department of Energy under
Grant No.~DE-FG02-96ER40956.

\end{acknowledgments}

\bibliographystyle{JHEP}
\bibliography{refs}

\providecommand{\href}[2]{#2}\begingroup\raggedright\begin{thebibliography}{10}

\bibitem{Ellis:2007ib}
S.~D. Ellis, J.~Huston, K.~Hatakeyama, P.~Loch, and M.~Tonnesmann, {\it {Jets
  in Hadron-Hadron Collisions}},  {\em Prog. Part. Nucl. Phys.} {\bf 60} (2008)
  484--551, \href{http://xxx.lanl.gov/abs/arXiv:0712.2447 [hep-ph]}{{\tt
  arXiv:0712.2447 [hep-ph]}}.

\bibitem{Shuryak}
E.~Shuryak, {\it Why does the quark gluon plasma at {RHIC} behave as a nearly
  ideal fluid?},  {\em Prog. Part. Nucl. Phys.} {\bf 53} (2004) 273--303,
  \href{http://xxx.lanl.gov/abs/hep-ph/0312227}{{\tt hep-ph/0312227}}.

\bibitem{Shuryak:2004cy}
E.~V. Shuryak, {\it What {RHIC} experiments and theory tell us about properties
  of quark-gluon plasma?},  {\em Nucl. Phys.} {\bf A750} (2005) 64--83,
  \href{http://xxx.lanl.gov/abs/hep-ph/0405066}{{\tt hep-ph/0405066}}.

\bibitem{Lin:2007fa}
S.~Lin and E.~Shuryak, {\it {Toward the AdS/CFT Gravity Dual for High Energy
  Collisions: II. The Stress Tensor on the Boundary}},
  \href{http://xxx.lanl.gov/abs/arXiv:0711.0736 [hep-th]}{{\tt arXiv:0711.0736
  [hep-th]}}.

\bibitem{Chernicoff:2008sa}
M.~Chernicoff and A.~Guijosa, {\it {Acceleration, Energy Loss and Screening in
  Strongly- Coupled Gauge Theories}},  {\em JHEP} {\bf 06} (2008) 005,
  \href{http://xxx.lanl.gov/abs/0803.3070}{{\tt 0803.3070}}.

\bibitem{Herzog:2006gh}
C.~P. Herzog, A.~Karch, P.~Kovtun, C.~Kozcaz, and L.~G. Yaffe, {\it {Energy
  loss of a heavy quark moving through N = 4 supersymmetric Yang-Mills
  plasma}},  {\em JHEP} {\bf 07} (2006) 013,
  \href{http://xxx.lanl.gov/abs/hep-th/0605158}{{\tt hep-th/0605158}}.

\bibitem{Gubser:2006bz}
S.~S. Gubser, {\it Drag force in {AdS/CFT}},  {\em Phys. Rev.} {\bf D74} (2006)
  126005, \href{http://xxx.lanl.gov/abs/hep-th/0605182}{{\tt hep-th/0605182}}.

\bibitem{Gubser:2008as}
S.~S. Gubser, D.~R. Gulotta, S.~S. Pufu, and F.~D. Rocha, {\it {Gluon energy
  loss in the gauge-string duality}},
  \href{http://xxx.lanl.gov/abs/arXiv:0803.1470 [hep-th]}{{\tt arXiv:0803.1470
  [hep-th]}}.

\bibitem{Hatta:2008tx}
Y.~Hatta, E.~Iancu, and A.~H. Mueller, {\it {Jet evolution in the N=4 SYM
  plasma at strong coupling}},  \href{http://xxx.lanl.gov/abs/0803.2481}{{\tt
  0803.2481}}.

\bibitem{Basham:1977iq}
C.~L. Basham, L.~S. Brown, S.~D. Ellis, and S.~T. Love, {\it {Electron -
  Positron Annihilation Energy Pattern in Quantum Chromodynamics:
  Asymptotically Free Perturbation Theory}},  {\em Phys. Rev.} {\bf D17} (1978)
  2298.

\bibitem{Basham:1978bw}
C.~L. Basham, L.~S. Brown, S.~D. Ellis, and S.~T. Love, {\it {Energy
  Correlations in electron - Positron Annihilation: Testing QCD}},  {\em Phys.
  Rev. Lett.} {\bf 41} (1978) 1585.

\bibitem{Basham:1978zq}
C.~L. Basham, L.~S. Brown, S.~D. Ellis, and S.~T. Love, {\it {Energy
  Correlations in electron-Positron Annihilation in Quantum Chromodynamics:
  Asymptotically Free Perturbation Theory}},  {\em Phys. Rev.} {\bf D19} (1979)
  2018.

\bibitem{Hofman:2008ar}
D.~M. Hofman and J.~Maldacena, {\it {Conformal collider physics: Energy and
  charge correlations}},  \href{http://xxx.lanl.gov/abs/arXiv:0803.1467
  [hep-th]}{{\tt arXiv:0803.1467 [hep-th]}}.

\bibitem{Polchinski:2002jw}
J.~Polchinski and M.~J. Strassler, {\it {Deep inelastic scattering and
  gauge/string duality}},  {\em JHEP} {\bf 05} (2003) 012,
  \href{http://xxx.lanl.gov/abs/hep-th/0209211}{{\tt hep-th/0209211}}.

\bibitem{Strassler:2008bv}
M.~J. Strassler, {\it {Why Unparticle Models with Mass Gaps are Examples of
  Hidden Valleys}},  \href{http://xxx.lanl.gov/abs/arXiv:0801.0629
  [hep-ph]}{{\tt arXiv:0801.0629 [hep-ph]}}.

\bibitem{Maldacena:1997re}
J.~M. Maldacena, {\it The large {$N$} limit of superconformal field theories
  and supergravity},  {\em Adv. Theor. Math. Phys.} {\bf 2} (1998) 231--252,
  \href{http://xxx.lanl.gov/abs/hep-th/9711200}{{\tt hep-th/9711200}}.

\bibitem{Gubser:1998bc}
S.~S. Gubser, I.~R. Klebanov, and A.~M. Polyakov, {\it Gauge theory correlators
  from non-critical string theory},  {\em Phys. Lett.} {\bf B428} (1998)
  105--114, \href{http://xxx.lanl.gov/abs/hep-th/9802109}{{\tt
  hep-th/9802109}}.

\bibitem{Witten:1998qj}
E.~Witten, {\it Anti-de {S}itter space and holography},  {\em Adv. Theor. Math.
  Phys.} {\bf 2} (1998) 253--291,
  \href{http://xxx.lanl.gov/abs/hep-th/9802150}{{\tt hep-th/9802150}}.

\bibitem{Maldacena:1998im}
J.~M. Maldacena, {\it {Wilson loops in large N field theories}},  {\em Phys.
  Rev. Lett.} {\bf 80} (1998) 4859--4862,
  \href{http://xxx.lanl.gov/abs/hep-th/9803002}{{\tt hep-th/9803002}}.

\bibitem{Rey:1998ik}
S.-J. Rey and J.-T. Yee, {\it {Macroscopic strings as heavy quarks in large N
  gauge theory and anti-de Sitter supergravity}},  {\em Eur. Phys. J.} {\bf
  C22} (2001) 379--394, \href{http://xxx.lanl.gov/abs/hep-th/9803001}{{\tt
  hep-th/9803001}}.

\bibitem{Gubser:2002tv}
S.~S. Gubser, I.~R. Klebanov, and A.~M. Polyakov, {\it {A semi-classical limit
  of the gauge/string correspondence}},  {\em Nucl. Phys.} {\bf B636} (2002)
  99--114, \href{http://xxx.lanl.gov/abs/hep-th/0204051}{{\tt hep-th/0204051}}.

\bibitem{Karch:2002sh}
A.~Karch and E.~Katz, {\it Adding flavor to {AdS/CFT}},  {\em JHEP} {\bf 06}
  (2002) 043, \href{http://xxx.lanl.gov/abs/hep-th/0205236}{{\tt
  hep-th/0205236}}.

\bibitem{Son:2002sd}
D.~T. Son and A.~O. Starinets, {\it Minkowski-space correlators in {AdS/CFT}
  correspondence: Recipe and applications},  {\em JHEP} {\bf 09} (2002) 042,
  \href{http://xxx.lanl.gov/abs/hep-th/0205051}{{\tt hep-th/0205051}}.

\bibitem{Ellis:1993ik}
S.~D. Ellis, {\it {Collider jets in perturbation theory}},
  \href{http://xxx.lanl.gov/abs/hep-ph/9306280}{{\tt hep-ph/9306280}}.

\bibitem{Seymour:1997kj}
M.~H. Seymour, {\it {Jet shapes in hadron collisions: Higher orders,
  resummation and hadronization}},  {\em Nucl. Phys.} {\bf B513} (1998)
  269--300, \href{http://xxx.lanl.gov/abs/hep-ph/9707338}{{\tt
  hep-ph/9707338}}.

\bibitem{Kovtun:2003wp}
P.~Kovtun, D.~T. Son, and A.~O. Starinets, {\it {Holography and hydrodynamics:
  Diffusion on stretched horizons}},  {\em JHEP} {\bf 10} (2003) 064,
  \href{http://xxx.lanl.gov/abs/hep-th/0309213}{{\tt hep-th/0309213}}.

\bibitem{Myers:2007we}
R.~C. Myers, A.~O. Starinets, and R.~M. Thomson, {\it {Holographic spectral
  functions and diffusion constants for fundamental matter}},  {\em JHEP} {\bf
  11} (2007) 091, \href{http://xxx.lanl.gov/abs/0706.0162}{{\tt 0706.0162}}.

\bibitem{Chesler:2007an}
P.~M. Chesler and L.~G. Yaffe, {\it The wake of a quark moving through a
  strongly-coupled {$\mathcal N=4$} supersymmetric {Yang-Mills} plasma},
  \href{http://xxx.lanl.gov/abs/arXiv:0706.0368 [hep-th]}{{\tt arXiv:0706.0368
  [hep-th]}}.

\bibitem{Chesler:2007sv}
P.~M. Chesler and L.~G. Yaffe, {\it {The stress-energy tensor of a quark moving
  through a strongly-coupled N=4 supersymmetric Yang-Mills plasma: comparing
  hydrodynamics and AdS/CFT}},  \href{http://xxx.lanl.gov/abs/0712.0050}{{\tt
  0712.0050}}.

\bibitem{Gubser:2008vz}
S.~S. Gubser and A.~Yarom, {\it {Linearized hydrodynamics from probe-sources in
  the gauge- string duality}},  \href{http://xxx.lanl.gov/abs/0803.0081}{{\tt
  0803.0081}}.

\bibitem{Kovtun:2005ev}
P.~K. Kovtun and A.~O. Starinets, {\it Quasinormal modes and holography},  {\em
  Phys. Rev.} {\bf D72} (2005) 086009,
  \href{http://xxx.lanl.gov/abs/hep-th/0506184}{{\tt hep-th/0506184}}.

\bibitem{Danielsson:1998wt}
U.~H. Danielsson, E.~Keski-Vakkuri, and M.~Kruczenski, {\it {Vacua,
  propagators, and holographic probes in AdS/CFT}},  {\em JHEP} {\bf 01} (1999)
  002, \href{http://xxx.lanl.gov/abs/hep-th/9812007}{{\tt hep-th/9812007}}.

\bibitem{Gubser:2007xz}
S.~S. Gubser, S.~S. Pufu, and A.~Yarom, {\it Energy disturbances due to a
  moving quark from gauge-string duality},  {\em JHEP} {\bf 09} (2007) 108,
  \href{http://xxx.lanl.gov/abs/arXiv:0706.0213 [hep-th]}{{\tt arXiv:0706.0213
  [hep-th]}}.

\bibitem{Gubser:2007ga}
S.~S. Gubser, S.~S. Pufu, and A.~Yarom, {\it Sonic booms and diffusion wakes
  generated by a heavy quark in thermal {AdS/CFT}},
  \href{http://xxx.lanl.gov/abs/arXiv: 0706.4307 [hep-th]}{{\tt arXiv:
  0706.4307 [hep-th]}}.

\bibitem{Gubser:2007zr}
S.~S. Gubser, S.~S. Pufu, and A.~Yarom, {\it {Shock waves from heavy-quark
  mesons in AdS/CFT}},  \href{http://xxx.lanl.gov/abs/0711.1415}{{\tt
  0711.1415}}.

\end{thebibliography}\endgroup

\end{document}